\documentclass[twocolumn,reprint, amsmath, amssymb, pra,aps, superscriptaddress,aps]{revtex4-2}

\pdfoutput=1
\usepackage{graphicx}
\usepackage{dcolumn}
\usepackage{bm}
\usepackage{xcolor}
\usepackage[normalem]{ulem}
\usepackage{amsmath}
\usepackage[english]{babel}
\usepackage{braket}
\usepackage{units}
\usepackage[T1]{fontenc}
\usepackage{comment}

\newcommand{\Hamiltonian}{\hat{\mathcal{H}}}

\makeatletter
\def\maketitle{
\@author@finish
\title@column\titleblock@produce
\suppressfloats[t]}
\makeatother

\preprint{APS/123-QED}

\begin{document}

\title{The waves-in-space Purcell effect for superconducting qubits}
\author{Param Patel}
\affiliation{Department of Physics and Astronomy, University of Pittsburgh, Pittsburgh, PA, USA}
\affiliation{Department of Applied Physics, Yale University, New Haven, CT, USA}
\author{Mingkang Xia}
\affiliation{Department of Physics and Astronomy, University of Pittsburgh, Pittsburgh, PA, USA}
\affiliation{Department of Applied Physics, Yale University, New Haven, CT, USA}
\author{Chao Zhou}
\affiliation{Department of Physics and Astronomy, University of Pittsburgh, Pittsburgh, PA, USA}
\author{Pinlei Lu}
\affiliation{Department of Physics and Astronomy, University of Pittsburgh, Pittsburgh, PA, USA}
\author{Xi Cao}
\affiliation{Department of Physics and Astronomy, University of Pittsburgh, Pittsburgh, PA, USA}
\author{Israa Yusuf}
\affiliation{Department of Physics and Astronomy, University of Pittsburgh, Pittsburgh, PA, USA}
\affiliation{Department of Applied Physics, Yale University, New Haven, CT, USA}
\author{Jacob Repicky}
\affiliation{Department of Physics and Astronomy, University of Pittsburgh, Pittsburgh, PA, USA}
\affiliation{Department of Applied Physics, Yale University, New Haven, CT, USA}
\author{Michael Hatridge}
\affiliation{Department of Physics and Astronomy, University of Pittsburgh, Pittsburgh, PA, USA}
\affiliation{Department of Applied Physics, Yale University, New Haven, CT, USA}
\date{\today}

\begin{abstract}
Quantum information processing, especially with quantum error correction, requires both long-lived qubits and fast, quantum non-demolition readout.  In superconducting circuits this leads to the requirement to both strongly couple qubits, such as transmons, to readout modes while also protecting them from associated Purcell decay through the readout port. So-called Purcell filters can provide this protection, at the cost of significant increases in circuit components and complexity.  However, as we demonstrate in this work, visualizing the qubit fields in space reveals locations where the qubit fields are strong and cavity fields weak; simply placing ports at these locations provides intrinsic Purcell protection. For a $\lambda/2$ readout mode in the `chip-in-tube' geometry, we show  both millisecond level Purcell protection and, conversely, greatly enhanced Purcell decay (qubit lifetime of 1~$\mu$s) simply by relocating the readout port. This method of integrating the Purcell protection into the qubit-cavity geometry can be generalized to other 3D implementations, such as post-cavities, as well as planar geometries.  For qubit frequencies below the readout mode this effect is quite distinct from the multi-mode Purcell effect, which we demonstrate in a 3D-post geometry where we show both Purcell protection of the qubit while spoiling the quality factor of higher cavity harmonics to protect against dephasing due to stray photons in these modes.
\end{abstract}

\maketitle

\section{Introduction}
Superconducting circuits \cite{DevoretSchoelkopfReview2013,krantz2019,Blais2021} whether targeted at near-term, Noisy Intermediate-Scale Quantum (NISQ) information processing \cite{GoogleQuantum2019,Wu2021,Zhu2022} or long-term, error-corrected machines, place a premium on the lifetime of individual qubits.  In circuit quantum electrodynamics (cQED) \cite{Blais2004,OliverReview2019} one key limit is the auxiliary linear resonator each qubit is coupled to, which both allows QND readout of the qubit state \cite{Braginsky1980} but also creates a channel through which the qubit can decay (the so-called Purcell effect \cite{Purcell1946}).  High-fidelity readout necessitates that a port connected to the outside environment be strongly coupled to the cavity ($\kappa_c \ge 1$~MHz),  with a dispersive shift of the same order \cite{Clerk2010,DevoretSchoelkopfReview2013,OliverReview2019}. 

\begin{figure}[b!]  
    \centering
	\includegraphics[]{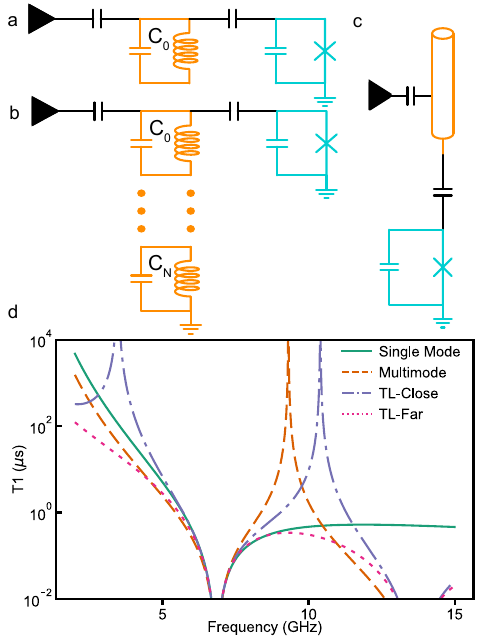}
	\caption[Microwave Model]{\textbf{Loss versus readout circuit model.} single-mode (a), multi-mode (b), and full transmission line (c) circuit models for a qubit and readout mode. d) Numerical simulations of the qubit lifetime using the different circuit models.  The full transmission line model contains a sweet spot below the lowest readout mode not explained by the single-mode or multi-mode models.}
	\label{fig:circuits}
\end{figure}
Almost all readout modes are made using transmission line resonators or three-dimensional cavities, which have an infinite series of higher harmonics.  This modifies the single-mode Purcell loss (Fig.~\ref{fig:circuits}a) \cite {Nakamura1999,GeerlingsThesis} depending on the qubit's frequency and physical placement \cite{Houck2008}. This `multi-mode Purcell effect' is often modeled by considering the infinitely many higher-order harmonics of the cavity as a series array of harmonic oscillators (Fig.~\ref{fig:circuits}b) \cite{BBQ2012}. As an example, we consider the case of a one-dimensional transmission line cavity with a qubit at one end (the most common configuration in 2D \cite{IBM2017, OliverReview2019,GoogleQuantum2019, Wang2022} and qubit-in-tube \cite{AxlineThesis,Ganjam2024} experiments). The most notable features captured by the circuit model in Fig.~\ref{fig:circuits}b for this configuration are sweet spots between pairs of modes where the qubit decay to each mode is equal and destructively interferes (see Fig.~\ref{fig:circuits}d at 9 GHz).  However, modeling the true system schematic (Fig.~\ref{fig:circuits}c) using Microwave Office or a 3D simulator such as HFSS, reveals an additional sweet spot below the lowest cavity mode (see Fig.~\ref{fig:circuits}d around 3 GHz). 

Many experiments place the qubit below the lowest cavity mode, and so these points have  been found in experiments where reported lifetimes are much greater than the single-mode Purcell limit, for example reference \cite {Sunada2022}.  However, the process of locating and explaining them is typically obscured by the multi-mode picture, whereas we will show that visualizing the fields is quite straightforward. Further, while Purcell decay can be suppressed by the use of auxiliary modes, i.e., Purcell filters to attenuate emission of qubit mode photons through the readout port \cite{Reed2010,Jeffrey2014,Bronn2015,Walter2017}, this technique increases the design complexity of each qubit's readout circuit and can be strengthened or replaced by proper readout port placement.
\begin{figure}[h!]  
    \centering
	\includegraphics[width = \linewidth]{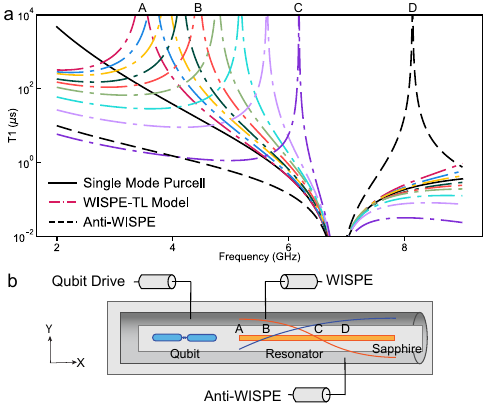}
	\caption[WISPE]{\textbf{The waves in space Purcell effect.} a) Through a numerical circuit simulation solver, different circuits were modeled to explore the nature of the ``sweet spot'' from the transmission line model. In solid black, we show the single-mode, lumped model. The dot-dashed lines correspond to different readout coupler positions along the length of the resonator closer to the qubit. Letters A-C correspond to the specific positions indicated in (b) on the top of the graph. Each curve in (a) corresponds to a unique port position along the resonator which has a unique range of qubit frequencies in its ``sweet spot'' giving rise to some geometric interference pattern between port position and the qubit frequency. The dashed, black line corresponds to a port position (D) on the far end of the resonator where the lifetime is increased above the lowest resonator mode, but remarkably lower than the single-mode limit below the resonator mode, which we dub the ``anti-WISPE'' point. b) A physical diagram of the qubit-resonator system in a 3D tube. Mapped are the ports used to drive and readout the system and the positions A-D are denoted along the length of the resonator that correspond to curves on the above plot. The blue and orange lines on the resonator correspond to the qubit and resonator electric fields, respectively.}
	\label{fig:wispe}
\end{figure}

In this paper, we explain the existence of this sweet spot, its frequency dependence, and spatial extent in transmission line resonators in both 2D and 3D using the concept of the waves-in-space Purcell Effect (WISPE). We demonstrate both millisecond-level Purcell protection and strong Purcell decay in the same qubit-resonator system, dependent only on the readout coupling port position in our 3D tube geometry. In each case, we consider the power needed to drive a $\pi$-pulse \cite{GeerlingsThesis} through the cavity port versus a weakly coupled qubit drive port. We show that the ratio of these powers predicts the level of Purcell protection and can even be used to extract the internal losses of the qubit mode, which may be of use for diagnosing mechanisms of qubit decay. Moreover, we demonstrate in a 3D post-cavity system that we can strongly modify the lifetimes of the higher harmonics of the cavity mode without degrading the $T_1$ of the qubit through the utilization of the spatial extent of the qubit/cavity modes. 

\section{Coherence calculations}

As a first example, we model the qubit-resonator system as a circuit with lumped components \cite{Koch2007}. We model the resonator mode as a parallel LC circuit ($\omega_r=7~GHz$) and the qubit, in our case the transmon, as a Josephson junction shunted by a capacitor, as shown in Fig.~\ref{fig:circuits}a \cite{Makhlin2001,Koch2007}. The couplings between the two objects are defined by the capacitors between the circuit components. We calculate the frequency response of this circuit using Microwave Office and treating the qubit as if it was a virtual port and assume that the resonator is coupled to a $50~\Omega$ resistive load as the environment. We monitor the lifetime of the qubit by evaluating $T_1=\frac{C}{Re[Y(\omega)]}$, where $Y(\omega)$ is the admittance seen from the qubit and $C$ is the qubit capacitance \cite{Houck2008}. As a result, we replicate the results of the atomic case following the equation  $\gamma_q = \kappa_r \left(\frac{g}{\Delta}\right)^2$, where $\kappa_r$ is the linewidth of the resonator mode, $g$ is the coupling strength between the qubit and resonator, and $\Delta$ is the frequency difference between the qubit and resonator \cite{Purcell1946}. Here, $\gamma_q$ is the decay rate of the qubit due to its coupling to the lumped resonator mode.

In the vast majority of modern systems, the cavity or resonator is not truly a lumped object. This necessitates consideration of the infinitely many higher-order harmonics of the fundamental resonator mode. To capture this in our simulations, we can replace the one, lumped resonator mode with many lumped resonator modes, a practice which has been used to create analogs of qubits and other superconducting circuits in other quantum simulation tools such as black-box quantization \cite{BBQ2012,Minev2021}. In our Microwave Office simulation, we truncate the infinite modes to include 3 resonator modes akin to the model shown in Figure~\ref{fig:circuits}b.  By including these additional resonator modes, we find unique increases in qubit lifetime between the resonator modes (subsequent harmonics of the resonator) corresponding to the true multi-mode Purcell effect, in which the higher-order modes destructively interfere and create frequency regimes where the lifetime of the qubit is greatly enhanced. However, these regions are above the lowest resonator mode and in cQED, the qubit is traditionally below the lowest resonator frequency\cite{AxlineThesis}.

If, instead, we model the transmission line circuit without approximating it as a series of modes, we capture its true behavior. In addition, in our microwave system we can easily modify the location of the readout port along the transmission line while holding its total length and frequency fixed. As shown in Figure~\ref{fig:wispe}a, for a given readout port location on the near side of the resonator, a qubit frequency range exists where a Purcell effect null/lifetime sweet spot can be found below the readout mode frequency.  We can model the system as a simple transmon qubit dispersively coupled to a stripline resonator suspended in a 3D superconducting tube. By plotting the electromagnetic fields at the qubit frequency (see SI~Fig.~\ref{fig:panflute}b,c), we find a region along the resonator where there is a null point in the qubit field and strong cavity fields, in direct analog to the simple circuit model. Thus, there is inherent protection within the geometry of the system. By changing the frequency of the qubit or the resonator, the null point moves along the half of the resonator closer to the qubit. As a consequence, for each qubit-resonator frequency pair, we can identify a unique position along the resonator where the qubit lifetime is enhanced. Due to the unique wave-like and spatial nature of this phenomenon, we dub this effect the waves-in-space Purcell effect (WISPE).

This leads to two further consequences. Visualizing the waves in space can be extended to more complex geometries like a qubit suspended in a 3D post-cavity where there can exist regions where the qubit fields are weak but the cavity fields are simultaneously strong (see SI~Fig.~\ref{fig:3Dpost}b,c). Another consequence is that we can alter the coupling of the higher order cavity modes to the qubit to limit photon shot noise dephasing \cite{SearsThesis}. By changing the qubit coupling to the higher order cavity modes, we would change the multi-mode Purcell loss. In our model, we show that by analyzing the waves in space, the qubit lifetime is agnostic to these multi-modal alterations we can make and we can benefit from techniques that could increase the $T_{2R}$ of our qubit. 

To demonstrate the power of our technique we have conducted the following experiments. First, we have utilized a simple coaxial tube geometry \cite{AxlineThesis} and a Ta transmon qubit \cite{Place2021} dispersively coupled to a stripline resonator to demonstrate strong Purcell protection by placing a readout port at the qubit null fields (which we dub the ``WISPE'' spot). In addition, we show the same qubit can be severely limited by Purcell loss, even lower than the single mode Purcell limit, by changing the readout port position to the worst possible location (which we dub the ``Anti-WISPE'' spot). Next, we extended WISPE to a 3D post-cavity and alter the higher order cavity modes through use of the innate cavity geometry and the introduction of Eccosorb into the cavity lid. We show that the lifetime of this qubit is unchanged with the introduction of Eccosorb, and, further, the qubit $T_2$ was enhanced which we attribute to decoupling it from dephasing due to stray photons in these modes \cite{SearsThesis}.

\section{Experiment Setup and Results}

\subsection{3D Tubes}
\begin{figure}[h]  
    \centering
	\includegraphics[width=\linewidth]{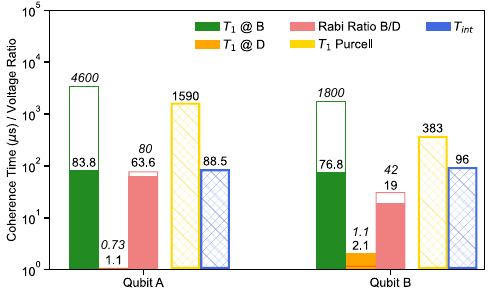}
	\caption[Tubes]{\textbf{WISPE vs. Anti-WISPE.} From subsequent measurements, qubit lifetimes and $\pi$-pulse voltage ratios are taken from different readout port configurations. Measured quantities of qubit coherence and this Rabi voltage ratio are displayed using solid bars on the bar chart. Hollow bars around these solid bars mark the simulated values (in italics) of these qubits in HFSS. The lifetime in the heavily Purcell limited case and the Rabi voltage ratios match closely to simulated values. Hashed bars display calculated values of Purcell loss and cumulative losses (calculation in SI Sec. \ref{ssec:coherencecalc}).}
	\label{fig:tube}
\end{figure}

Our experiment is designed to demonstrate the principle of WISPE by measuring both long and short coherence times for a single transmon simply by placing coupling ports at different locations along the readout mode.  A diagram of the two coupling ports is shown in SI~Figure~\ref{fig:panflute}a. One readout coupling port is located at the null point of the qubit electric field for a qubit at $\omega_{ge}/2 \pi = 5.05$~GHz labeled the ``WISPE'' port and another is located at the worst available placement which we label the ``anti-WISPE'' port. We use spark-plug connectors which can be demounted and replaced without changing the qubit/resonator chip or even their positioning within the tube. We also varied the height of the coupling port to hold the cavity linewidth $\kappa/2 \pi = 5 $~MHz constant at the two locations.  For two qubits, we performed comparisons between two runs, one each in which the cavity ports were connected in the WISPE and anti-WISPE configurations. However, we can do more to elucidate the internal losses of the qubit by comparing the $\pi$-pulse areas used to drive a qubit transition between the $\ket{g}-\ket{e}$ states from the fixed qubit port and the two readout ports, providing insight on the relative couplings of the qubit to each port. The data on qubit $T_1$ and the $\pi$-pulse ratios from the qubit/readout ports is tracked on the first three solid bars in Figure~\ref{fig:tube}. Simulated values of these qubits from HFSS are plotted in hollow bars with their values in italics around the solid bars. We note that the measured Purcell-limited qubit coherence and the Rabi voltage ratio from the ports have good correlation to simulated values.

Using the pulse-derived relative couplings and the changes in lifetime between both readout configurations, we extract the qubit loss rate due to Purcell decay to the protected WISPE port and the cumulative internal losses of the qubit.  This calculation is detailed in SI~Sec.~\ref{ssec:coherencecalc}, and relies on the assumptions that internal losses do not change between cooldowns and are long compared to the anti-WISPE decay rate. We could, in principle, use a single port configuration to infer the Purcell loss, however, as our qubit port is both very sensitive to fine placement details, and, by design, very weak, we find this two-configuration comparison using the qubit port as a benchmark to be far more robust. The calculated loss rates are displayed in Figure~\ref{fig:tube} as the hashed bars.

Qubit lifetimes of approximately 80 $\mu s$ were achieved in the WISPE configuration for both qubits. When the coupling port was swapped to the `anti-WISPE' configuration, the lifetime dropped to 1 $\mu s$, showing, even in the presence of significant internal losses, at least an 80-fold modification in the Purcell effect by only changing the port placement. However, the true effect is much larger, as the $\pi$-pulse ratio of the two WISPE ports is 63.6 for qubit A and 19 for qubit B, from which we calculate a Purcell decay time of 1.59 and 0.383 ms, respectively, demonstrating that we can achieve practical Purcell protection at the millisecond level solely from careful placement of the cavity port!  The simulated values in HFSS are higher, with qubit A's higher degree of protection due primarily to its closer match to the design frequency (see SI~Sec.~\ref{ssec:housings}).  We attribute the remaining discrepancies between predicted and actual Purcell protection to the fine details of the port placement, which for Purcell quality factors in the $10^7$ to $10^8$ range are quite sensitive. We believe that these qubits could have 100 ms level Purcell protection if the qubits had better matched the sweet spot frequency.

\subsection{3D Cavity}

\begin{figure}[]  
    \centering
	\includegraphics[width=\linewidth]{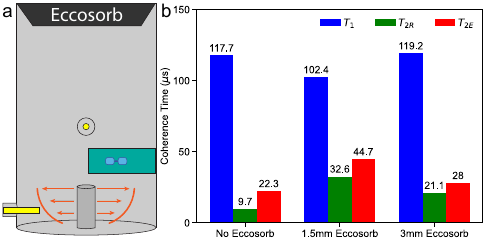}
	\caption[3D cavity]{\textbf{3D Cavity} a) Extending WISPE to a post-cavity geometry allows us to introduce Eccosorb into the system. We can try to increase the $T_{2R}$ of the qubit by introducing more loss into the higher order cavity modes without altering the lifetime of the qubit. By doing so and removing the coupling of the qubit to higher harmonics of the cavity, we also effectively demonstrate how WISPE is purely a geometric phenomena. b) Using more and more Eccosorb in the lid of the post-cavity, we measured the same qubit across subsequent runs. We note how the $T_1$ is largely unchanged, but there are slight upward trends with $T_2$ with the introduction of Eccosorb.}
	\label{fig:cavity}
\end{figure}

We extend the use of WISPE for guiding port placement from a tube system to a more complex geometry: a 3D coaxial post-cavity \cite{Reagor2016}, a schematic of which is shown in Fig.~\ref{fig:cavity}a. By visualizing the electric fields of the qubit and cavity at their respective frequencies, we look for regions where the qubit fields are weak in regions where the cavity fields are strong. Effectively, we are maximizing $\left|\frac{\vec{E_q}\cdot \vec{E_c}}{\vec{E_c}\cdot \vec{E_c}}\right|$ over a region large enough to place a readout coupling port (see SI~Fig.~\ref{fig:3Dpost}). The port was placed using this figure of merit to identify a region, in this case in the shadow of the qubit fields created by the post and far from the qubit, rather than quite near the qubit in the previous geometry, where there is strong Purcell protection.

 More, for our 3D post cavity, we have designed the cutoff of the filter segment \cite{Reagor2016} so that all modes but the fundamental occupy the full height of the cavity. We inserted Eccosorb \cite{Laird}, a lossy rf material, into the lid of the cavity to reduce the coupling of all the resonator modes except for the lowest mode. This heavily reduces the ability of higher modes to dephase our qubit due to thermal occupancy \cite{SearsThesis} without degrading the qubit and fundamental mode's lifetimes, in agreement with finite element simulations shown in SI~Sec.~\ref{ssec:highermodes}. 

We demonstrated this via a series of measurements where the lifetime of a qubit was measured in 3 different cavity configurations. We first measured the qubit with no Eccosorb, then with 1.5 mm of Eccosorb, and then with 3 mm of Eccosorb adhered to the cavity lid (Fig.~\ref{fig:cavity}b). Overall, the $T_1$ of the qubit was largely unchanged between cooldowns and the measured $T_{2R}$ times increased by a factor of roughly two for the Eccosorb configurations, indicating suppression of dephasing due to stray photons. This demonstrates that the multi-mode Purcell effect is not applicable in the case where the qubit is below the lowest cavity mode since by actively removing the higher order cavity modes, we have tampered with the multi-mode interference seen in Fig.~\ref{fig:circuits}d. This is further evidence that the spatial phenomena of WISPE is distinct from the multi-mode Purcell effect.

\section{Conclusion} 

In conclusion, WISPE and visualization of the electric fields of both the qubit and resonator at their respective frequencies provides an excellent design tool to guide the placement of the readout coupling port in any superconducting qubit architecture. We have demonstrated this effect in $\lambda/2$ stripline resonators, mirroring the readout modes in large-scale 2D transmon  architectures \cite{GoogleQuantum2019,IBM2017}, where large system sizes put a premium on avoiding the space and complexity of external Purcell filters. We have also demonstrated WISPE in a more complicated geometry,  the 3D post-cavity. More, for $T_1$ characterization, the ability to vastly change the Purcell effect for individual samples can provide a more robust and accurate tool for separating internal and external losses, guiding efforts to improve qubit materials and lifetimes.

\bibliography{refs}

\begin{thebibliography}{32}%
\makeatletter
\providecommand \@ifxundefined [1]{%
 \@ifx{#1\undefined}
}%
\providecommand \@ifnum [1]{%
 \ifnum #1\expandafter \@firstoftwo
 \else \expandafter \@secondoftwo
 \fi
}%
\providecommand \@ifx [1]{%
 \ifx #1\expandafter \@firstoftwo
 \else \expandafter \@secondoftwo
 \fi
}%
\providecommand \natexlab [1]{#1}%
\providecommand \enquote  [1]{``#1''}%
\providecommand \bibnamefont  [1]{#1}%
\providecommand \bibfnamefont [1]{#1}%
\providecommand \citenamefont [1]{#1}%
\providecommand \href@noop [0]{\@secondoftwo}%
\providecommand \href [0]{\begingroup \@sanitize@url \@href}%
\providecommand \@href[1]{\@@startlink{#1}\@@href}%
\providecommand \@@href[1]{\endgroup#1\@@endlink}%
\providecommand \@sanitize@url [0]{\catcode `\\12\catcode `\$12\catcode
  `\&12\catcode `\#12\catcode `\^12\catcode `\_12\catcode `\%12\relax}%
\providecommand \@@startlink[1]{}%
\providecommand \@@endlink[0]{}%
\providecommand \url  [0]{\begingroup\@sanitize@url \@url }%
\providecommand \@url [1]{\endgroup\@href {#1}{\urlprefix }}%
\providecommand \urlprefix  [0]{URL }%
\providecommand \Eprint [0]{\href }%
\providecommand \doibase [0]{https://doi.org/}%
\providecommand \selectlanguage [0]{\@gobble}%
\providecommand \bibinfo  [0]{\@secondoftwo}%
\providecommand \bibfield  [0]{\@secondoftwo}%
\providecommand \translation [1]{[#1]}%
\providecommand \BibitemOpen [0]{}%
\providecommand \bibitemStop [0]{}%
\providecommand \bibitemNoStop [0]{.\EOS\space}%
\providecommand \EOS [0]{\spacefactor3000\relax}%
\providecommand \BibitemShut  [1]{\csname bibitem#1\endcsname}%
\let\auto@bib@innerbib\@empty
\bibitem [{\citenamefont {Devoret}\ and\ \citenamefont
  {Schoelkopf}(2013)}]{DevoretSchoelkopfReview2013}%
  \BibitemOpen
  \bibfield  {author} {\bibinfo {author} {\bibfnamefont {M.~H.}\ \bibnamefont
  {Devoret}}\ and\ \bibinfo {author} {\bibfnamefont {R.~J.}\ \bibnamefont
  {Schoelkopf}},\ }\bibfield  {title} {\bibinfo {title} {Superconducting
  circuits for quantum information: An outlook},\ }\href
  {https://doi.org/10.1126/science.1231930} {\bibfield  {journal} {\bibinfo
  {journal} {Science}\ }\textbf {\bibinfo {volume} {339}},\ \bibinfo {pages}
  {1169} (\bibinfo {year} {2013})},\ \Eprint
  {https://arxiv.org/abs/https://www.science.org/doi/pdf/10.1126/science.1231930}
  {https://www.science.org/doi/pdf/10.1126/science.1231930} \BibitemShut
  {NoStop}%
\bibitem [{\citenamefont {Krantz}\ \emph
  {et~al.}(2019{\natexlab{a}})\citenamefont {Krantz}, \citenamefont
  {Kjaergaard}, \citenamefont {Yan}, \citenamefont {Orlando}, \citenamefont
  {Gustavsson},\ and\ \citenamefont {Oliver}}]{krantz2019}%
  \BibitemOpen
  \bibfield  {author} {\bibinfo {author} {\bibfnamefont {P.}~\bibnamefont
  {Krantz}}, \bibinfo {author} {\bibfnamefont {M.}~\bibnamefont {Kjaergaard}},
  \bibinfo {author} {\bibfnamefont {F.}~\bibnamefont {Yan}}, \bibinfo {author}
  {\bibfnamefont {T.~P.}\ \bibnamefont {Orlando}}, \bibinfo {author}
  {\bibfnamefont {S.}~\bibnamefont {Gustavsson}},\ and\ \bibinfo {author}
  {\bibfnamefont {W.~D.}\ \bibnamefont {Oliver}},\ }\bibfield  {title}
  {\bibinfo {title} {A quantum engineer's guide to superconducting qubits},\
  }\href {https://doi.org/10.1063/1.5089550} {\bibfield  {journal} {\bibinfo
  {journal} {Applied Physics Reviews}\ }\textbf {\bibinfo {volume} {6}},\
  \bibinfo {pages} {021318} (\bibinfo {year} {2019}{\natexlab{a}})}\BibitemShut
  {NoStop}%
\bibitem [{\citenamefont {Blais}\ \emph {et~al.}(2021)\citenamefont {Blais},
  \citenamefont {Grimsmo}, \citenamefont {Girvin},\ and\ \citenamefont
  {Wallraff}}]{Blais2021}%
  \BibitemOpen
  \bibfield  {author} {\bibinfo {author} {\bibfnamefont {A.}~\bibnamefont
  {Blais}}, \bibinfo {author} {\bibfnamefont {A.~L.}\ \bibnamefont {Grimsmo}},
  \bibinfo {author} {\bibfnamefont {S.~M.}\ \bibnamefont {Girvin}},\ and\
  \bibinfo {author} {\bibfnamefont {A.}~\bibnamefont {Wallraff}},\ }\bibfield
  {title} {\bibinfo {title} {Circuit quantum electrodynamics},\ }\href
  {https://doi.org/10.1103/RevModPhys.93.025005} {\bibfield  {journal}
  {\bibinfo  {journal} {Rev. Mod. Phys.}\ }\textbf {\bibinfo {volume} {93}},\
  \bibinfo {pages} {025005} (\bibinfo {year} {2021})}\BibitemShut {NoStop}%
\bibitem [{\citenamefont {Arute}\ \emph {et~al.}(2019)\citenamefont {Arute},
  \citenamefont {Arya}, \citenamefont {Babbush}, \citenamefont {Bacon},
  \citenamefont {Bardin}, \citenamefont {Barends}, \citenamefont {Biswas},
  \citenamefont {Boixo}, \citenamefont {Brandao}, \citenamefont {Buell} \emph
  {et~al.}}]{GoogleQuantum2019}%
  \BibitemOpen
  \bibfield  {author} {\bibinfo {author} {\bibfnamefont {F.}~\bibnamefont
  {Arute}}, \bibinfo {author} {\bibfnamefont {K.}~\bibnamefont {Arya}},
  \bibinfo {author} {\bibfnamefont {R.}~\bibnamefont {Babbush}}, \bibinfo
  {author} {\bibfnamefont {D.}~\bibnamefont {Bacon}}, \bibinfo {author}
  {\bibfnamefont {J.}~\bibnamefont {Bardin}}, \bibinfo {author} {\bibfnamefont
  {R.}~\bibnamefont {Barends}}, \bibinfo {author} {\bibfnamefont
  {R.}~\bibnamefont {Biswas}}, \bibinfo {author} {\bibfnamefont
  {S.}~\bibnamefont {Boixo}}, \bibinfo {author} {\bibfnamefont
  {F.}~\bibnamefont {Brandao}}, \bibinfo {author} {\bibfnamefont
  {D.}~\bibnamefont {Buell}}, \emph {et~al.},\ }\bibfield  {title} {\bibinfo
  {title} {Quantum supremacy using a programmable superconducting processor.
  2019},\ }\href@noop {} {\bibfield  {journal} {\bibinfo  {journal} {Nature}\ }
  (\bibinfo {year} {2019})}\BibitemShut {NoStop}%
\bibitem [{\citenamefont {Wu}\ \emph {et~al.}(2021)\citenamefont {Wu},
  \citenamefont {Bao}, \citenamefont {Cao}, \citenamefont {Chen}, \citenamefont
  {Chen}, \citenamefont {Chen}, \citenamefont {Chung}, \citenamefont {Deng},
  \citenamefont {Du}, \citenamefont {Fan}, \citenamefont {Gong}, \citenamefont
  {Guo}, \citenamefont {Guo}, \citenamefont {Guo}, \citenamefont {Han},
  \citenamefont {Hong}, \citenamefont {Huang}, \citenamefont {Huo},
  \citenamefont {Li}, \citenamefont {Li}, \citenamefont {Li}, \citenamefont
  {Li}, \citenamefont {Liang}, \citenamefont {Lin}, \citenamefont {Lin},
  \citenamefont {Qian}, \citenamefont {Qiao}, \citenamefont {Rong},
  \citenamefont {Su}, \citenamefont {Sun}, \citenamefont {Wang}, \citenamefont
  {Wang}, \citenamefont {Wu}, \citenamefont {Xu}, \citenamefont {Yan},
  \citenamefont {Yang}, \citenamefont {Yang}, \citenamefont {Ye}, \citenamefont
  {Yin}, \citenamefont {Ying}, \citenamefont {Yu}, \citenamefont {Zha},
  \citenamefont {Zhang}, \citenamefont {Zhang}, \citenamefont {Zhang},
  \citenamefont {Zhang}, \citenamefont {Zhao}, \citenamefont {Zhao},
  \citenamefont {Zhou}, \citenamefont {Zhu}, \citenamefont {Lu}, \citenamefont
  {Peng}, \citenamefont {Zhu},\ and\ \citenamefont {Pan}}]{Wu2021}%
  \BibitemOpen
  \bibfield  {author} {\bibinfo {author} {\bibfnamefont {Y.}~\bibnamefont
  {Wu}}, \bibinfo {author} {\bibfnamefont {W.-S.}\ \bibnamefont {Bao}},
  \bibinfo {author} {\bibfnamefont {S.}~\bibnamefont {Cao}}, \bibinfo {author}
  {\bibfnamefont {F.}~\bibnamefont {Chen}}, \bibinfo {author} {\bibfnamefont
  {M.-C.}\ \bibnamefont {Chen}}, \bibinfo {author} {\bibfnamefont
  {X.}~\bibnamefont {Chen}}, \bibinfo {author} {\bibfnamefont {T.-H.}\
  \bibnamefont {Chung}}, \bibinfo {author} {\bibfnamefont {H.}~\bibnamefont
  {Deng}}, \bibinfo {author} {\bibfnamefont {Y.}~\bibnamefont {Du}}, \bibinfo
  {author} {\bibfnamefont {D.}~\bibnamefont {Fan}}, \bibinfo {author}
  {\bibfnamefont {M.}~\bibnamefont {Gong}}, \bibinfo {author} {\bibfnamefont
  {C.}~\bibnamefont {Guo}}, \bibinfo {author} {\bibfnamefont {C.}~\bibnamefont
  {Guo}}, \bibinfo {author} {\bibfnamefont {S.}~\bibnamefont {Guo}}, \bibinfo
  {author} {\bibfnamefont {L.}~\bibnamefont {Han}}, \bibinfo {author}
  {\bibfnamefont {L.}~\bibnamefont {Hong}}, \bibinfo {author} {\bibfnamefont
  {H.-L.}\ \bibnamefont {Huang}}, \bibinfo {author} {\bibfnamefont {Y.-H.}\
  \bibnamefont {Huo}}, \bibinfo {author} {\bibfnamefont {L.}~\bibnamefont
  {Li}}, \bibinfo {author} {\bibfnamefont {N.}~\bibnamefont {Li}}, \bibinfo
  {author} {\bibfnamefont {S.}~\bibnamefont {Li}}, \bibinfo {author}
  {\bibfnamefont {Y.}~\bibnamefont {Li}}, \bibinfo {author} {\bibfnamefont
  {F.}~\bibnamefont {Liang}}, \bibinfo {author} {\bibfnamefont
  {C.}~\bibnamefont {Lin}}, \bibinfo {author} {\bibfnamefont {J.}~\bibnamefont
  {Lin}}, \bibinfo {author} {\bibfnamefont {H.}~\bibnamefont {Qian}}, \bibinfo
  {author} {\bibfnamefont {D.}~\bibnamefont {Qiao}}, \bibinfo {author}
  {\bibfnamefont {H.}~\bibnamefont {Rong}}, \bibinfo {author} {\bibfnamefont
  {H.}~\bibnamefont {Su}}, \bibinfo {author} {\bibfnamefont {L.}~\bibnamefont
  {Sun}}, \bibinfo {author} {\bibfnamefont {L.}~\bibnamefont {Wang}}, \bibinfo
  {author} {\bibfnamefont {S.}~\bibnamefont {Wang}}, \bibinfo {author}
  {\bibfnamefont {D.}~\bibnamefont {Wu}}, \bibinfo {author} {\bibfnamefont
  {Y.}~\bibnamefont {Xu}}, \bibinfo {author} {\bibfnamefont {K.}~\bibnamefont
  {Yan}}, \bibinfo {author} {\bibfnamefont {W.}~\bibnamefont {Yang}}, \bibinfo
  {author} {\bibfnamefont {Y.}~\bibnamefont {Yang}}, \bibinfo {author}
  {\bibfnamefont {Y.}~\bibnamefont {Ye}}, \bibinfo {author} {\bibfnamefont
  {J.}~\bibnamefont {Yin}}, \bibinfo {author} {\bibfnamefont {C.}~\bibnamefont
  {Ying}}, \bibinfo {author} {\bibfnamefont {J.}~\bibnamefont {Yu}}, \bibinfo
  {author} {\bibfnamefont {C.}~\bibnamefont {Zha}}, \bibinfo {author}
  {\bibfnamefont {C.}~\bibnamefont {Zhang}}, \bibinfo {author} {\bibfnamefont
  {H.}~\bibnamefont {Zhang}}, \bibinfo {author} {\bibfnamefont
  {K.}~\bibnamefont {Zhang}}, \bibinfo {author} {\bibfnamefont
  {Y.}~\bibnamefont {Zhang}}, \bibinfo {author} {\bibfnamefont
  {H.}~\bibnamefont {Zhao}}, \bibinfo {author} {\bibfnamefont {Y.}~\bibnamefont
  {Zhao}}, \bibinfo {author} {\bibfnamefont {L.}~\bibnamefont {Zhou}}, \bibinfo
  {author} {\bibfnamefont {Q.}~\bibnamefont {Zhu}}, \bibinfo {author}
  {\bibfnamefont {C.-Y.}\ \bibnamefont {Lu}}, \bibinfo {author} {\bibfnamefont
  {C.-Z.}\ \bibnamefont {Peng}}, \bibinfo {author} {\bibfnamefont
  {X.}~\bibnamefont {Zhu}},\ and\ \bibinfo {author} {\bibfnamefont {J.-W.}\
  \bibnamefont {Pan}},\ }\bibfield  {title} {\bibinfo {title} {Strong quantum
  computational advantage using a superconducting quantum processor},\ }\href
  {https://doi.org/10.1103/PhysRevLett.127.180501} {\bibfield  {journal}
  {\bibinfo  {journal} {Phys. Rev. Lett.}\ }\textbf {\bibinfo {volume} {127}},\
  \bibinfo {pages} {180501} (\bibinfo {year} {2021})}\BibitemShut {NoStop}%
\bibitem [{\citenamefont {Zhu}\ \emph {et~al.}(2022)\citenamefont {Zhu},
  \citenamefont {Cao}, \citenamefont {Chen}, \citenamefont {Chen},
  \citenamefont {Chen}, \citenamefont {Chung}, \citenamefont {Deng},
  \citenamefont {Du}, \citenamefont {Fan}, \citenamefont {Gong}, \citenamefont
  {Guo}, \citenamefont {Guo}, \citenamefont {Guo}, \citenamefont {Han},
  \citenamefont {Hong}, \citenamefont {Huang}, \citenamefont {Huo},
  \citenamefont {Li}, \citenamefont {Li}, \citenamefont {Li}, \citenamefont
  {Li}, \citenamefont {Liang}, \citenamefont {Lin}, \citenamefont {Lin},
  \citenamefont {Qian}, \citenamefont {Qiao}, \citenamefont {Rong},
  \citenamefont {Su}, \citenamefont {Sun}, \citenamefont {Wang}, \citenamefont
  {Wang}, \citenamefont {Wu}, \citenamefont {Wu}, \citenamefont {Xu},
  \citenamefont {Yan}, \citenamefont {Yang}, \citenamefont {Yang},
  \citenamefont {Ye}, \citenamefont {Yin}, \citenamefont {Ying}, \citenamefont
  {Yu}, \citenamefont {Zha}, \citenamefont {Zhang}, \citenamefont {Zhang},
  \citenamefont {Zhang}, \citenamefont {Zhang}, \citenamefont {Zhao},
  \citenamefont {Zhao}, \citenamefont {Zhou}, \citenamefont {Lu}, \citenamefont
  {Peng}, \citenamefont {Zhu},\ and\ \citenamefont {Pan}}]{Zhu2022}%
  \BibitemOpen
  \bibfield  {author} {\bibinfo {author} {\bibfnamefont {Q.}~\bibnamefont
  {Zhu}}, \bibinfo {author} {\bibfnamefont {S.}~\bibnamefont {Cao}}, \bibinfo
  {author} {\bibfnamefont {F.}~\bibnamefont {Chen}}, \bibinfo {author}
  {\bibfnamefont {M.-C.}\ \bibnamefont {Chen}}, \bibinfo {author}
  {\bibfnamefont {X.}~\bibnamefont {Chen}}, \bibinfo {author} {\bibfnamefont
  {T.-H.}\ \bibnamefont {Chung}}, \bibinfo {author} {\bibfnamefont
  {H.}~\bibnamefont {Deng}}, \bibinfo {author} {\bibfnamefont {Y.}~\bibnamefont
  {Du}}, \bibinfo {author} {\bibfnamefont {D.}~\bibnamefont {Fan}}, \bibinfo
  {author} {\bibfnamefont {M.}~\bibnamefont {Gong}}, \bibinfo {author}
  {\bibfnamefont {C.}~\bibnamefont {Guo}}, \bibinfo {author} {\bibfnamefont
  {C.}~\bibnamefont {Guo}}, \bibinfo {author} {\bibfnamefont {S.}~\bibnamefont
  {Guo}}, \bibinfo {author} {\bibfnamefont {L.}~\bibnamefont {Han}}, \bibinfo
  {author} {\bibfnamefont {L.}~\bibnamefont {Hong}}, \bibinfo {author}
  {\bibfnamefont {H.-L.}\ \bibnamefont {Huang}}, \bibinfo {author}
  {\bibfnamefont {Y.-H.}\ \bibnamefont {Huo}}, \bibinfo {author} {\bibfnamefont
  {L.}~\bibnamefont {Li}}, \bibinfo {author} {\bibfnamefont {N.}~\bibnamefont
  {Li}}, \bibinfo {author} {\bibfnamefont {S.}~\bibnamefont {Li}}, \bibinfo
  {author} {\bibfnamefont {Y.}~\bibnamefont {Li}}, \bibinfo {author}
  {\bibfnamefont {F.}~\bibnamefont {Liang}}, \bibinfo {author} {\bibfnamefont
  {C.}~\bibnamefont {Lin}}, \bibinfo {author} {\bibfnamefont {J.}~\bibnamefont
  {Lin}}, \bibinfo {author} {\bibfnamefont {H.}~\bibnamefont {Qian}}, \bibinfo
  {author} {\bibfnamefont {D.}~\bibnamefont {Qiao}}, \bibinfo {author}
  {\bibfnamefont {H.}~\bibnamefont {Rong}}, \bibinfo {author} {\bibfnamefont
  {H.}~\bibnamefont {Su}}, \bibinfo {author} {\bibfnamefont {L.}~\bibnamefont
  {Sun}}, \bibinfo {author} {\bibfnamefont {L.}~\bibnamefont {Wang}}, \bibinfo
  {author} {\bibfnamefont {S.}~\bibnamefont {Wang}}, \bibinfo {author}
  {\bibfnamefont {D.}~\bibnamefont {Wu}}, \bibinfo {author} {\bibfnamefont
  {Y.}~\bibnamefont {Wu}}, \bibinfo {author} {\bibfnamefont {Y.}~\bibnamefont
  {Xu}}, \bibinfo {author} {\bibfnamefont {K.}~\bibnamefont {Yan}}, \bibinfo
  {author} {\bibfnamefont {W.}~\bibnamefont {Yang}}, \bibinfo {author}
  {\bibfnamefont {Y.}~\bibnamefont {Yang}}, \bibinfo {author} {\bibfnamefont
  {Y.}~\bibnamefont {Ye}}, \bibinfo {author} {\bibfnamefont {J.}~\bibnamefont
  {Yin}}, \bibinfo {author} {\bibfnamefont {C.}~\bibnamefont {Ying}}, \bibinfo
  {author} {\bibfnamefont {J.}~\bibnamefont {Yu}}, \bibinfo {author}
  {\bibfnamefont {C.}~\bibnamefont {Zha}}, \bibinfo {author} {\bibfnamefont
  {C.}~\bibnamefont {Zhang}}, \bibinfo {author} {\bibfnamefont
  {H.}~\bibnamefont {Zhang}}, \bibinfo {author} {\bibfnamefont
  {K.}~\bibnamefont {Zhang}}, \bibinfo {author} {\bibfnamefont
  {Y.}~\bibnamefont {Zhang}}, \bibinfo {author} {\bibfnamefont
  {H.}~\bibnamefont {Zhao}}, \bibinfo {author} {\bibfnamefont {Y.}~\bibnamefont
  {Zhao}}, \bibinfo {author} {\bibfnamefont {L.}~\bibnamefont {Zhou}}, \bibinfo
  {author} {\bibfnamefont {C.-Y.}\ \bibnamefont {Lu}}, \bibinfo {author}
  {\bibfnamefont {C.-Z.}\ \bibnamefont {Peng}}, \bibinfo {author}
  {\bibfnamefont {X.}~\bibnamefont {Zhu}},\ and\ \bibinfo {author}
  {\bibfnamefont {J.-W.}\ \bibnamefont {Pan}},\ }\bibfield  {title} {\bibinfo
  {title} {Quantum computational advantage via 60-qubit 24-cycle random circuit
  sampling},\ }\href
  {https://doi.org/https://doi.org/10.1016/j.scib.2021.10.017} {\bibfield
  {journal} {\bibinfo  {journal} {Science Bulletin}\ }\textbf {\bibinfo
  {volume} {67}},\ \bibinfo {pages} {240} (\bibinfo {year} {2022})}\BibitemShut
  {NoStop}%
\bibitem [{\citenamefont {Blais}\ \emph {et~al.}(2004)\citenamefont {Blais},
  \citenamefont {Huang}, \citenamefont {Wallraff}, \citenamefont {Girvin},\
  and\ \citenamefont {Schoelkopf}}]{Blais2004}%
  \BibitemOpen
  \bibfield  {author} {\bibinfo {author} {\bibfnamefont {A.}~\bibnamefont
  {Blais}}, \bibinfo {author} {\bibfnamefont {R.-S.}\ \bibnamefont {Huang}},
  \bibinfo {author} {\bibfnamefont {A.}~\bibnamefont {Wallraff}}, \bibinfo
  {author} {\bibfnamefont {S.~M.}\ \bibnamefont {Girvin}},\ and\ \bibinfo
  {author} {\bibfnamefont {R.~J.}\ \bibnamefont {Schoelkopf}},\ }\bibfield
  {title} {\bibinfo {title} {Cavity quantum electrodynamics for superconducting
  electrical circuits: An architecture for quantum computation},\ }\href
  {https://doi.org/10.1103/PhysRevA.69.062320} {\bibfield  {journal} {\bibinfo
  {journal} {Phys. Rev. A}\ }\textbf {\bibinfo {volume} {69}},\ \bibinfo
  {pages} {062320} (\bibinfo {year} {2004})}\BibitemShut {NoStop}%
\bibitem [{\citenamefont {Krantz}\ \emph
  {et~al.}(2019{\natexlab{b}})\citenamefont {Krantz}, \citenamefont
  {Kjaergaard}, \citenamefont {Yan}, \citenamefont {Orlando}, \citenamefont
  {Gustavsson},\ and\ \citenamefont {Oliver}}]{OliverReview2019}%
  \BibitemOpen
  \bibfield  {author} {\bibinfo {author} {\bibfnamefont {P.}~\bibnamefont
  {Krantz}}, \bibinfo {author} {\bibfnamefont {M.}~\bibnamefont {Kjaergaard}},
  \bibinfo {author} {\bibfnamefont {F.}~\bibnamefont {Yan}}, \bibinfo {author}
  {\bibfnamefont {T.~P.}\ \bibnamefont {Orlando}}, \bibinfo {author}
  {\bibfnamefont {S.}~\bibnamefont {Gustavsson}},\ and\ \bibinfo {author}
  {\bibfnamefont {W.~D.}\ \bibnamefont {Oliver}},\ }\bibfield  {title}
  {\bibinfo {title} {A quantum engineer's guide to superconducting qubits},\
  }\bibfield  {journal} {\bibinfo  {journal} {APPLIED PHYSICS REVIEWS}\
  }\textbf {\bibinfo {volume} {6}},\ \href {https://doi.org/10.1063/1.5089550}
  {10.1063/1.5089550} (\bibinfo {year} {2019}{\natexlab{b}})\BibitemShut
  {NoStop}%
\bibitem [{\citenamefont {Braginsky}\ \emph {et~al.}(1980)\citenamefont
  {Braginsky}, \citenamefont {Vorontsov},\ and\ \citenamefont
  {Thorne}}]{Braginsky1980}%
  \BibitemOpen
  \bibfield  {author} {\bibinfo {author} {\bibfnamefont {V.~B.}\ \bibnamefont
  {Braginsky}}, \bibinfo {author} {\bibfnamefont {Y.~I.}\ \bibnamefont
  {Vorontsov}},\ and\ \bibinfo {author} {\bibfnamefont {K.~S.}\ \bibnamefont
  {Thorne}},\ }\bibfield  {title} {\bibinfo {title} {Quantum nondemolition
  measurements},\ }\href {https://doi.org/10.1126/science.209.4456.547}
  {\bibfield  {journal} {\bibinfo  {journal} {Science}\ }\textbf {\bibinfo
  {volume} {209}},\ \bibinfo {pages} {547} (\bibinfo {year}
  {1980})}\BibitemShut {NoStop}%
\bibitem [{\citenamefont {Purcell}\ \emph {et~al.}(1946)\citenamefont
  {Purcell}, \citenamefont {Torrey},\ and\ \citenamefont
  {Pound}}]{Purcell1946}%
  \BibitemOpen
  \bibfield  {author} {\bibinfo {author} {\bibfnamefont {E.~M.}\ \bibnamefont
  {Purcell}}, \bibinfo {author} {\bibfnamefont {H.~C.}\ \bibnamefont
  {Torrey}},\ and\ \bibinfo {author} {\bibfnamefont {R.~V.}\ \bibnamefont
  {Pound}},\ }\bibfield  {title} {\bibinfo {title} {Resonance absorption by
  nuclear magnetic moments in a solid},\ }\href
  {https://doi.org/10.1103/PhysRev.69.37} {\bibfield  {journal} {\bibinfo
  {journal} {Phys. Rev.}\ }\textbf {\bibinfo {volume} {69}},\ \bibinfo {pages}
  {37} (\bibinfo {year} {1946})}\BibitemShut {NoStop}%
\bibitem [{\citenamefont {Clerk}\ \emph {et~al.}(2010)\citenamefont {Clerk},
  \citenamefont {Devoret}, \citenamefont {Girvin}, \citenamefont {Marquardt},\
  and\ \citenamefont {Schoelkopf}}]{Clerk2010}%
  \BibitemOpen
  \bibfield  {author} {\bibinfo {author} {\bibfnamefont {A.~A.}\ \bibnamefont
  {Clerk}}, \bibinfo {author} {\bibfnamefont {M.~H.}\ \bibnamefont {Devoret}},
  \bibinfo {author} {\bibfnamefont {S.~M.}\ \bibnamefont {Girvin}}, \bibinfo
  {author} {\bibfnamefont {F.}~\bibnamefont {Marquardt}},\ and\ \bibinfo
  {author} {\bibfnamefont {R.~J.}\ \bibnamefont {Schoelkopf}},\ }\bibfield
  {title} {\bibinfo {title} {Introduction to quantum noise, measurement, and
  amplification},\ }\href {https://doi.org/10.1103/RevModPhys.82.1155}
  {\bibfield  {journal} {\bibinfo  {journal} {Rev. Mod. Phys.}\ }\textbf
  {\bibinfo {volume} {82}},\ \bibinfo {pages} {1155} (\bibinfo {year}
  {2010})}\BibitemShut {NoStop}%
\bibitem [{\citenamefont {Nakamura}\ \emph {et~al.}(1999)\citenamefont
  {Nakamura}, \citenamefont {Pashkin},\ and\ \citenamefont
  {Tsai}}]{Nakamura1999}%
  \BibitemOpen
  \bibfield  {author} {\bibinfo {author} {\bibfnamefont {Y.}~\bibnamefont
  {Nakamura}}, \bibinfo {author} {\bibfnamefont {Y.~A.}\ \bibnamefont
  {Pashkin}},\ and\ \bibinfo {author} {\bibfnamefont {J.~S.}\ \bibnamefont
  {Tsai}},\ }\bibfield  {title} {\bibinfo {title} {Coherent control of
  macroscopic quantum states in a single-{Cooper}-pair box},\ }\href
  {https://doi.org/10.1038/19718} {\bibfield  {journal} {\bibinfo  {journal}
  {Nature}\ }\textbf {\bibinfo {volume} {398}},\ \bibinfo {pages} {786}
  (\bibinfo {year} {1999})},\ \bibinfo {note} {publisher: Nature Publishing
  Group}\BibitemShut {NoStop}%
\bibitem [{\citenamefont {Geerlings}(2013)}]{GeerlingsThesis}%
  \BibitemOpen
  \bibfield  {author} {\bibinfo {author} {\bibfnamefont {K.~L.}\ \bibnamefont
  {Geerlings}},\ }\emph {\bibinfo {title} {Improving Coherence of
  Superconducting Qubits and Resonators}},\ \href@noop {} {Ph.D. thesis},\
  \bibinfo  {school} {Yale University} (\bibinfo {year} {2013})\BibitemShut
  {NoStop}%
\bibitem [{\citenamefont {Houck}\ \emph {et~al.}(2008)\citenamefont {Houck},
  \citenamefont {Schreier}, \citenamefont {Johnson}, \citenamefont {Chow},
  \citenamefont {Koch}, \citenamefont {Gambetta}, \citenamefont {Schuster},
  \citenamefont {Frunzio}, \citenamefont {Devoret}, \citenamefont {Girvin},\
  and\ \citenamefont {Schoelkopf}}]{Houck2008}%
  \BibitemOpen
  \bibfield  {author} {\bibinfo {author} {\bibfnamefont {A.~A.}\ \bibnamefont
  {Houck}}, \bibinfo {author} {\bibfnamefont {J.~A.}\ \bibnamefont {Schreier}},
  \bibinfo {author} {\bibfnamefont {B.~R.}\ \bibnamefont {Johnson}}, \bibinfo
  {author} {\bibfnamefont {J.~M.}\ \bibnamefont {Chow}}, \bibinfo {author}
  {\bibfnamefont {J.}~\bibnamefont {Koch}}, \bibinfo {author} {\bibfnamefont
  {J.~M.}\ \bibnamefont {Gambetta}}, \bibinfo {author} {\bibfnamefont {D.~I.}\
  \bibnamefont {Schuster}}, \bibinfo {author} {\bibfnamefont {L.}~\bibnamefont
  {Frunzio}}, \bibinfo {author} {\bibfnamefont {M.~H.}\ \bibnamefont
  {Devoret}}, \bibinfo {author} {\bibfnamefont {S.~M.}\ \bibnamefont
  {Girvin}},\ and\ \bibinfo {author} {\bibfnamefont {R.~J.}\ \bibnamefont
  {Schoelkopf}},\ }\bibfield  {title} {\bibinfo {title} {Controlling the
  spontaneous emission of a superconducting transmon qubit},\ }\href
  {https://doi.org/10.1103/PhysRevLett.101.080502} {\bibfield  {journal}
  {\bibinfo  {journal} {Phys. Rev. Lett.}\ }\textbf {\bibinfo {volume} {101}},\
  \bibinfo {pages} {080502} (\bibinfo {year} {2008})}\BibitemShut {NoStop}%
\bibitem [{\citenamefont {Nigg}\ \emph {et~al.}(2012)\citenamefont {Nigg},
  \citenamefont {Paik}, \citenamefont {Vlastakis}, \citenamefont {Kirchmair},
  \citenamefont {Shankar}, \citenamefont {Frunzio}, \citenamefont {Devoret},
  \citenamefont {Schoelkopf},\ and\ \citenamefont {Girvin}}]{BBQ2012}%
  \BibitemOpen
  \bibfield  {author} {\bibinfo {author} {\bibfnamefont {S.~E.}\ \bibnamefont
  {Nigg}}, \bibinfo {author} {\bibfnamefont {H.}~\bibnamefont {Paik}}, \bibinfo
  {author} {\bibfnamefont {B.}~\bibnamefont {Vlastakis}}, \bibinfo {author}
  {\bibfnamefont {G.}~\bibnamefont {Kirchmair}}, \bibinfo {author}
  {\bibfnamefont {S.}~\bibnamefont {Shankar}}, \bibinfo {author} {\bibfnamefont
  {L.}~\bibnamefont {Frunzio}}, \bibinfo {author} {\bibfnamefont {M.~H.}\
  \bibnamefont {Devoret}}, \bibinfo {author} {\bibfnamefont {R.~J.}\
  \bibnamefont {Schoelkopf}},\ and\ \bibinfo {author} {\bibfnamefont {S.~M.}\
  \bibnamefont {Girvin}},\ }\bibfield  {title} {\bibinfo {title} {Black-box
  superconducting circuit quantization},\ }\href
  {https://doi.org/10.1103/PhysRevLett.108.240502} {\bibfield  {journal}
  {\bibinfo  {journal} {Phys. Rev. Lett.}\ }\textbf {\bibinfo {volume} {108}},\
  \bibinfo {pages} {240502} (\bibinfo {year} {2012})}\BibitemShut {NoStop}%
\bibitem [{\citenamefont {Ristè}\ \emph {et~al.}(2017)\citenamefont {Ristè},
  \citenamefont {da~Silva}, \citenamefont {Ryan}, \citenamefont {Cross},
  \citenamefont {Córcoles}, \citenamefont {Smolin}, \citenamefont {Gambetta},
  \citenamefont {Chow},\ and\ \citenamefont {Johnson}}]{IBM2017}%
  \BibitemOpen
  \bibfield  {author} {\bibinfo {author} {\bibfnamefont {D.}~\bibnamefont
  {Ristè}}, \bibinfo {author} {\bibfnamefont {M.~P.}\ \bibnamefont
  {da~Silva}}, \bibinfo {author} {\bibfnamefont {C.~A.}\ \bibnamefont {Ryan}},
  \bibinfo {author} {\bibfnamefont {A.~W.}\ \bibnamefont {Cross}}, \bibinfo
  {author} {\bibfnamefont {A.~D.}\ \bibnamefont {Córcoles}}, \bibinfo {author}
  {\bibfnamefont {J.~A.}\ \bibnamefont {Smolin}}, \bibinfo {author}
  {\bibfnamefont {J.~M.}\ \bibnamefont {Gambetta}}, \bibinfo {author}
  {\bibfnamefont {J.~M.}\ \bibnamefont {Chow}},\ and\ \bibinfo {author}
  {\bibfnamefont {B.~R.}\ \bibnamefont {Johnson}},\ }\bibfield  {title}
  {\bibinfo {title} {Demonstration of quantum advantage in machine learning},\
  }\bibfield  {journal} {\bibinfo  {journal} {npj Quantum Information}\
  }\textbf {\bibinfo {volume} {3}},\ \href
  {https://doi.org/10.1038/s41534-017-0017-3} {10.1038/s41534-017-0017-3}
  (\bibinfo {year} {2017})\BibitemShut {NoStop}%
\bibitem [{\citenamefont {Wang}\ \emph {et~al.}(2022)\citenamefont {Wang},
  \citenamefont {Li}, \citenamefont {Xu}, \citenamefont {Li}, \citenamefont
  {Wang}, \citenamefont {Yang}, \citenamefont {Mi}, \citenamefont {Liang},
  \citenamefont {Su}, \citenamefont {Yang}, \citenamefont {Wang}, \citenamefont
  {Wang}, \citenamefont {Li}, \citenamefont {Chen}, \citenamefont {Li},
  \citenamefont {Linghu}, \citenamefont {Han}, \citenamefont {Zhang},
  \citenamefont {Feng}, \citenamefont {Song}, \citenamefont {Ma}, \citenamefont
  {Zhang}, \citenamefont {Wang}, \citenamefont {Zhao}, \citenamefont {Liu},
  \citenamefont {Xue}, \citenamefont {Jin},\ and\ \citenamefont
  {Yu}}]{Wang2022}%
  \BibitemOpen
  \bibfield  {author} {\bibinfo {author} {\bibfnamefont {C.}~\bibnamefont
  {Wang}}, \bibinfo {author} {\bibfnamefont {X.}~\bibnamefont {Li}}, \bibinfo
  {author} {\bibfnamefont {H.}~\bibnamefont {Xu}}, \bibinfo {author}
  {\bibfnamefont {Z.}~\bibnamefont {Li}}, \bibinfo {author} {\bibfnamefont
  {J.}~\bibnamefont {Wang}}, \bibinfo {author} {\bibfnamefont {Z.}~\bibnamefont
  {Yang}}, \bibinfo {author} {\bibfnamefont {Z.}~\bibnamefont {Mi}}, \bibinfo
  {author} {\bibfnamefont {X.}~\bibnamefont {Liang}}, \bibinfo {author}
  {\bibfnamefont {T.}~\bibnamefont {Su}}, \bibinfo {author} {\bibfnamefont
  {C.}~\bibnamefont {Yang}}, \bibinfo {author} {\bibfnamefont {G.}~\bibnamefont
  {Wang}}, \bibinfo {author} {\bibfnamefont {W.}~\bibnamefont {Wang}}, \bibinfo
  {author} {\bibfnamefont {Y.}~\bibnamefont {Li}}, \bibinfo {author}
  {\bibfnamefont {M.}~\bibnamefont {Chen}}, \bibinfo {author} {\bibfnamefont
  {C.}~\bibnamefont {Li}}, \bibinfo {author} {\bibfnamefont {K.}~\bibnamefont
  {Linghu}}, \bibinfo {author} {\bibfnamefont {J.}~\bibnamefont {Han}},
  \bibinfo {author} {\bibfnamefont {Y.}~\bibnamefont {Zhang}}, \bibinfo
  {author} {\bibfnamefont {Y.}~\bibnamefont {Feng}}, \bibinfo {author}
  {\bibfnamefont {Y.}~\bibnamefont {Song}}, \bibinfo {author} {\bibfnamefont
  {T.}~\bibnamefont {Ma}}, \bibinfo {author} {\bibfnamefont {J.}~\bibnamefont
  {Zhang}}, \bibinfo {author} {\bibfnamefont {R.}~\bibnamefont {Wang}},
  \bibinfo {author} {\bibfnamefont {P.}~\bibnamefont {Zhao}}, \bibinfo {author}
  {\bibfnamefont {W.}~\bibnamefont {Liu}}, \bibinfo {author} {\bibfnamefont
  {G.}~\bibnamefont {Xue}}, \bibinfo {author} {\bibfnamefont {Y.}~\bibnamefont
  {Jin}},\ and\ \bibinfo {author} {\bibfnamefont {H.}~\bibnamefont {Yu}},\
  }\bibfield  {title} {\bibinfo {title} {Towards practical quantum computers:
  transmon qubit with a lifetime approaching 0.5 milliseconds},\ }\bibfield
  {journal} {\bibinfo  {journal} {npj Quantum Information}\ }\textbf {\bibinfo
  {volume} {8}},\ \href {https://doi.org/10.1038/s41534-021-00510-2}
  {10.1038/s41534-021-00510-2} (\bibinfo {year} {2022})\BibitemShut {NoStop}%
\bibitem [{\citenamefont {Axline}(2018)}]{AxlineThesis}%
  \BibitemOpen
  \bibfield  {author} {\bibinfo {author} {\bibfnamefont {C.~J.}\ \bibnamefont
  {Axline}},\ }\emph {\bibinfo {title} {Building Blocks for Modular Circuit QED
  Quantum Computing}},\ \href@noop {} {Ph.D. thesis},\ \bibinfo  {school} {Yale
  University} (\bibinfo {year} {2018})\BibitemShut {NoStop}%
\bibitem [{\citenamefont {Ganjam}\ \emph {et~al.}(2024)\citenamefont {Ganjam},
  \citenamefont {Wang}, \citenamefont {Lu}, \citenamefont {Banerjee},
  \citenamefont {Lei}, \citenamefont {Krayzman}, \citenamefont {Kisslinger},
  \citenamefont {Zhou}, \citenamefont {Li}, \citenamefont {Jia}, \citenamefont
  {Liu}, \citenamefont {Frunzio},\ and\ \citenamefont
  {Schoelkopf}}]{Ganjam2024}%
  \BibitemOpen
  \bibfield  {author} {\bibinfo {author} {\bibfnamefont {S.}~\bibnamefont
  {Ganjam}}, \bibinfo {author} {\bibfnamefont {Y.}~\bibnamefont {Wang}},
  \bibinfo {author} {\bibfnamefont {Y.}~\bibnamefont {Lu}}, \bibinfo {author}
  {\bibfnamefont {A.}~\bibnamefont {Banerjee}}, \bibinfo {author}
  {\bibfnamefont {C.~U.}\ \bibnamefont {Lei}}, \bibinfo {author} {\bibfnamefont
  {L.}~\bibnamefont {Krayzman}}, \bibinfo {author} {\bibfnamefont
  {K.}~\bibnamefont {Kisslinger}}, \bibinfo {author} {\bibfnamefont
  {C.}~\bibnamefont {Zhou}}, \bibinfo {author} {\bibfnamefont {R.}~\bibnamefont
  {Li}}, \bibinfo {author} {\bibfnamefont {Y.}~\bibnamefont {Jia}}, \bibinfo
  {author} {\bibfnamefont {M.}~\bibnamefont {Liu}}, \bibinfo {author}
  {\bibfnamefont {L.}~\bibnamefont {Frunzio}},\ and\ \bibinfo {author}
  {\bibfnamefont {R.~J.}\ \bibnamefont {Schoelkopf}},\ }\bibfield  {title}
  {\bibinfo {title} {Surpassing millisecond coherence in on chip
  superconducting quantum memories by optimizing materials and circuit
  design},\ }\bibfield  {journal} {\bibinfo  {journal} {Nature Communications}\
  }\textbf {\bibinfo {volume} {15}},\ \href
  {https://doi.org/10.1038/s41467-024-47857-6} {10.1038/s41467-024-47857-6}
  (\bibinfo {year} {2024})\BibitemShut {NoStop}%
\bibitem [{\citenamefont {Sunada}\ \emph {et~al.}(2022)\citenamefont {Sunada},
  \citenamefont {Kono}, \citenamefont {Ilves}, \citenamefont {Tamate},
  \citenamefont {Sugiyama}, \citenamefont {Tabuchi},\ and\ \citenamefont
  {Nakamura}}]{Sunada2022}%
  \BibitemOpen
  \bibfield  {author} {\bibinfo {author} {\bibfnamefont {Y.}~\bibnamefont
  {Sunada}}, \bibinfo {author} {\bibfnamefont {S.}~\bibnamefont {Kono}},
  \bibinfo {author} {\bibfnamefont {J.}~\bibnamefont {Ilves}}, \bibinfo
  {author} {\bibfnamefont {S.}~\bibnamefont {Tamate}}, \bibinfo {author}
  {\bibfnamefont {T.}~\bibnamefont {Sugiyama}}, \bibinfo {author}
  {\bibfnamefont {Y.}~\bibnamefont {Tabuchi}},\ and\ \bibinfo {author}
  {\bibfnamefont {Y.}~\bibnamefont {Nakamura}},\ }\bibfield  {title} {\bibinfo
  {title} {Fast readout and reset of a superconducting qubit coupled to a
  resonator with an intrinsic purcell filter},\ }\href
  {https://doi.org/10.1103/PhysRevApplied.17.044016} {\bibfield  {journal}
  {\bibinfo  {journal} {Phys. Rev. Appl.}\ }\textbf {\bibinfo {volume} {17}},\
  \bibinfo {pages} {044016} (\bibinfo {year} {2022})}\BibitemShut {NoStop}%
\bibitem [{\citenamefont {Reed}\ \emph {et~al.}(2010)\citenamefont {Reed},
  \citenamefont {Johnson}, \citenamefont {Houck}, \citenamefont {DiCarlo},
  \citenamefont {Chow}, \citenamefont {Schuster}, \citenamefont {Frunzio},\
  and\ \citenamefont {Schoelkopf}}]{Reed2010}%
  \BibitemOpen
  \bibfield  {author} {\bibinfo {author} {\bibfnamefont {M.~D.}\ \bibnamefont
  {Reed}}, \bibinfo {author} {\bibfnamefont {B.~R.}\ \bibnamefont {Johnson}},
  \bibinfo {author} {\bibfnamefont {A.~A.}\ \bibnamefont {Houck}}, \bibinfo
  {author} {\bibfnamefont {L.}~\bibnamefont {DiCarlo}}, \bibinfo {author}
  {\bibfnamefont {J.~M.}\ \bibnamefont {Chow}}, \bibinfo {author}
  {\bibfnamefont {D.~I.}\ \bibnamefont {Schuster}}, \bibinfo {author}
  {\bibfnamefont {L.}~\bibnamefont {Frunzio}},\ and\ \bibinfo {author}
  {\bibfnamefont {R.~J.}\ \bibnamefont {Schoelkopf}},\ }\bibfield  {title}
  {\bibinfo {title} {{Fast reset and suppressing spontaneous emission of a
  superconducting qubit}},\ }\href {https://doi.org/10.1063/1.3435463}
  {\bibfield  {journal} {\bibinfo  {journal} {Applied Physics Letters}\
  }\textbf {\bibinfo {volume} {96}},\ \bibinfo {pages} {203110} (\bibinfo
  {year} {2010})}\BibitemShut {NoStop}%
\bibitem [{\citenamefont {Jeffrey}\ \emph {et~al.}(2014)\citenamefont
  {Jeffrey}, \citenamefont {Sank}, \citenamefont {Mutus}, \citenamefont
  {White}, \citenamefont {Kelly}, \citenamefont {Barends}, \citenamefont
  {Chen}, \citenamefont {Chen}, \citenamefont {Chiaro}, \citenamefont
  {Dunsworth}, \citenamefont {Megrant}, \citenamefont {O'Malley}, \citenamefont
  {Neill}, \citenamefont {Roushan}, \citenamefont {Vainsencher}, \citenamefont
  {Wenner}, \citenamefont {Cleland},\ and\ \citenamefont
  {Martinis}}]{Jeffrey2014}%
  \BibitemOpen
  \bibfield  {author} {\bibinfo {author} {\bibfnamefont {E.}~\bibnamefont
  {Jeffrey}}, \bibinfo {author} {\bibfnamefont {D.}~\bibnamefont {Sank}},
  \bibinfo {author} {\bibfnamefont {J.~Y.}\ \bibnamefont {Mutus}}, \bibinfo
  {author} {\bibfnamefont {T.~C.}\ \bibnamefont {White}}, \bibinfo {author}
  {\bibfnamefont {J.}~\bibnamefont {Kelly}}, \bibinfo {author} {\bibfnamefont
  {R.}~\bibnamefont {Barends}}, \bibinfo {author} {\bibfnamefont
  {Y.}~\bibnamefont {Chen}}, \bibinfo {author} {\bibfnamefont {Z.}~\bibnamefont
  {Chen}}, \bibinfo {author} {\bibfnamefont {B.}~\bibnamefont {Chiaro}},
  \bibinfo {author} {\bibfnamefont {A.}~\bibnamefont {Dunsworth}}, \bibinfo
  {author} {\bibfnamefont {A.}~\bibnamefont {Megrant}}, \bibinfo {author}
  {\bibfnamefont {P.~J.~J.}\ \bibnamefont {O'Malley}}, \bibinfo {author}
  {\bibfnamefont {C.}~\bibnamefont {Neill}}, \bibinfo {author} {\bibfnamefont
  {P.}~\bibnamefont {Roushan}}, \bibinfo {author} {\bibfnamefont
  {A.}~\bibnamefont {Vainsencher}}, \bibinfo {author} {\bibfnamefont
  {J.}~\bibnamefont {Wenner}}, \bibinfo {author} {\bibfnamefont {A.~N.}\
  \bibnamefont {Cleland}},\ and\ \bibinfo {author} {\bibfnamefont {J.~M.}\
  \bibnamefont {Martinis}},\ }\bibfield  {title} {\bibinfo {title} {Fast
  accurate state measurement with superconducting qubits},\ }\href
  {https://doi.org/10.1103/PhysRevLett.112.190504} {\bibfield  {journal}
  {\bibinfo  {journal} {Phys. Rev. Lett.}\ }\textbf {\bibinfo {volume} {112}},\
  \bibinfo {pages} {190504} (\bibinfo {year} {2014})}\BibitemShut {NoStop}%
\bibitem [{\citenamefont {Bronn}\ \emph {et~al.}(2015)\citenamefont {Bronn},
  \citenamefont {Liu}, \citenamefont {Hertzberg}, \citenamefont {Córcoles},
  \citenamefont {Houck}, \citenamefont {Gambetta},\ and\ \citenamefont
  {Chow}}]{Bronn2015}%
  \BibitemOpen
  \bibfield  {author} {\bibinfo {author} {\bibfnamefont {N.~T.}\ \bibnamefont
  {Bronn}}, \bibinfo {author} {\bibfnamefont {Y.}~\bibnamefont {Liu}}, \bibinfo
  {author} {\bibfnamefont {J.~B.}\ \bibnamefont {Hertzberg}}, \bibinfo {author}
  {\bibfnamefont {A.~D.}\ \bibnamefont {Córcoles}}, \bibinfo {author}
  {\bibfnamefont {A.~A.}\ \bibnamefont {Houck}}, \bibinfo {author}
  {\bibfnamefont {J.~M.}\ \bibnamefont {Gambetta}},\ and\ \bibinfo {author}
  {\bibfnamefont {J.~M.}\ \bibnamefont {Chow}},\ }\bibfield  {title} {\bibinfo
  {title} {{Broadband filters for abatement of spontaneous emission in circuit
  quantum electrodynamics}},\ }\href {https://doi.org/10.1063/1.4934867}
  {\bibfield  {journal} {\bibinfo  {journal} {Applied Physics Letters}\
  }\textbf {\bibinfo {volume} {107}},\ \bibinfo {pages} {172601} (\bibinfo
  {year} {2015})}\BibitemShut {NoStop}%
\bibitem [{\citenamefont {Walter}\ \emph {et~al.}(2017)\citenamefont {Walter},
  \citenamefont {Kurpiers}, \citenamefont {Gasparinetti}, \citenamefont
  {Magnard}, \citenamefont {Poto\ifmmode~\check{c}\else \v{c}\fi{}nik},
  \citenamefont {Salath\'e}, \citenamefont {Pechal}, \citenamefont {Mondal},
  \citenamefont {Oppliger}, \citenamefont {Eichler},\ and\ \citenamefont
  {Wallraff}}]{Walter2017}%
  \BibitemOpen
  \bibfield  {author} {\bibinfo {author} {\bibfnamefont {T.}~\bibnamefont
  {Walter}}, \bibinfo {author} {\bibfnamefont {P.}~\bibnamefont {Kurpiers}},
  \bibinfo {author} {\bibfnamefont {S.}~\bibnamefont {Gasparinetti}}, \bibinfo
  {author} {\bibfnamefont {P.}~\bibnamefont {Magnard}}, \bibinfo {author}
  {\bibfnamefont {A.}~\bibnamefont {Poto\ifmmode~\check{c}\else
  \v{c}\fi{}nik}}, \bibinfo {author} {\bibfnamefont {Y.}~\bibnamefont
  {Salath\'e}}, \bibinfo {author} {\bibfnamefont {M.}~\bibnamefont {Pechal}},
  \bibinfo {author} {\bibfnamefont {M.}~\bibnamefont {Mondal}}, \bibinfo
  {author} {\bibfnamefont {M.}~\bibnamefont {Oppliger}}, \bibinfo {author}
  {\bibfnamefont {C.}~\bibnamefont {Eichler}},\ and\ \bibinfo {author}
  {\bibfnamefont {A.}~\bibnamefont {Wallraff}},\ }\bibfield  {title} {\bibinfo
  {title} {Rapid high-fidelity single-shot dispersive readout of
  superconducting qubits},\ }\href
  {https://doi.org/10.1103/PhysRevApplied.7.054020} {\bibfield  {journal}
  {\bibinfo  {journal} {Phys. Rev. Appl.}\ }\textbf {\bibinfo {volume} {7}},\
  \bibinfo {pages} {054020} (\bibinfo {year} {2017})}\BibitemShut {NoStop}%
\bibitem [{\citenamefont {Koch}\ \emph {et~al.}(2007)\citenamefont {Koch},
  \citenamefont {Yu}, \citenamefont {Gambetta}, \citenamefont {Houck},
  \citenamefont {Schuster}, \citenamefont {Majer}, \citenamefont {Blais},
  \citenamefont {Devoret}, \citenamefont {Girvin},\ and\ \citenamefont
  {Schoelkopf}}]{Koch2007}%
  \BibitemOpen
  \bibfield  {author} {\bibinfo {author} {\bibfnamefont {J.}~\bibnamefont
  {Koch}}, \bibinfo {author} {\bibfnamefont {T.~M.}\ \bibnamefont {Yu}},
  \bibinfo {author} {\bibfnamefont {J.}~\bibnamefont {Gambetta}}, \bibinfo
  {author} {\bibfnamefont {A.~A.}\ \bibnamefont {Houck}}, \bibinfo {author}
  {\bibfnamefont {D.~I.}\ \bibnamefont {Schuster}}, \bibinfo {author}
  {\bibfnamefont {J.}~\bibnamefont {Majer}}, \bibinfo {author} {\bibfnamefont
  {A.}~\bibnamefont {Blais}}, \bibinfo {author} {\bibfnamefont {M.~H.}\
  \bibnamefont {Devoret}}, \bibinfo {author} {\bibfnamefont {S.~M.}\
  \bibnamefont {Girvin}},\ and\ \bibinfo {author} {\bibfnamefont {R.~J.}\
  \bibnamefont {Schoelkopf}},\ }\bibfield  {title} {\bibinfo {title}
  {Charge-insensitive qubit design derived from the cooper pair box},\ }\href
  {https://doi.org/10.1103/PhysRevA.76.042319} {\bibfield  {journal} {\bibinfo
  {journal} {Phys. Rev. A}\ }\textbf {\bibinfo {volume} {76}},\ \bibinfo
  {pages} {042319} (\bibinfo {year} {2007})}\BibitemShut {NoStop}%
\bibitem [{\citenamefont {Makhlin}\ \emph {et~al.}(2001)\citenamefont
  {Makhlin}, \citenamefont {Sch\"on},\ and\ \citenamefont
  {Shnirman}}]{Makhlin2001}%
  \BibitemOpen
  \bibfield  {author} {\bibinfo {author} {\bibfnamefont {Y.}~\bibnamefont
  {Makhlin}}, \bibinfo {author} {\bibfnamefont {G.}~\bibnamefont {Sch\"on}},\
  and\ \bibinfo {author} {\bibfnamefont {A.}~\bibnamefont {Shnirman}},\
  }\bibfield  {title} {\bibinfo {title} {Quantum-state engineering with
  josephson-junction devices},\ }\href
  {https://doi.org/10.1103/RevModPhys.73.357} {\bibfield  {journal} {\bibinfo
  {journal} {Rev. Mod. Phys.}\ }\textbf {\bibinfo {volume} {73}},\ \bibinfo
  {pages} {357} (\bibinfo {year} {2001})}\BibitemShut {NoStop}%
\bibitem [{\citenamefont {Minev}\ \emph {et~al.}(2021)\citenamefont {Minev},
  \citenamefont {Leghtas}, \citenamefont {Mundhada}, \citenamefont
  {Christakis}, \citenamefont {Pop},\ and\ \citenamefont
  {Devoret}}]{Minev2021}%
  \BibitemOpen
  \bibfield  {author} {\bibinfo {author} {\bibfnamefont {Z.~K.}\ \bibnamefont
  {Minev}}, \bibinfo {author} {\bibfnamefont {Z.}~\bibnamefont {Leghtas}},
  \bibinfo {author} {\bibfnamefont {S.~O.}\ \bibnamefont {Mundhada}}, \bibinfo
  {author} {\bibfnamefont {L.}~\bibnamefont {Christakis}}, \bibinfo {author}
  {\bibfnamefont {I.~M.}\ \bibnamefont {Pop}},\ and\ \bibinfo {author}
  {\bibfnamefont {M.~H.}\ \bibnamefont {Devoret}},\ }\bibfield  {title}
  {\bibinfo {title} {Energy-participation quantization of josephson circuits},\
  }\bibfield  {journal} {\bibinfo  {journal} {npj Quantum Information}\
  }\textbf {\bibinfo {volume} {7}},\ \href
  {https://doi.org/10.1038/s41534-021-00461-8} {10.1038/s41534-021-00461-8}
  (\bibinfo {year} {2021})\BibitemShut {NoStop}%
\bibitem [{\citenamefont {Sears}(2013)}]{SearsThesis}%
  \BibitemOpen
  \bibfield  {author} {\bibinfo {author} {\bibfnamefont {A.~P.}\ \bibnamefont
  {Sears}},\ }\emph {\bibinfo {title} {Extending COherence in Superconducting
  Qubits: from microseconds to milliseconds}},\ \href@noop {} {Ph.D. thesis},\
  \bibinfo  {school} {Yale University} (\bibinfo {year} {2013})\BibitemShut
  {NoStop}%
\bibitem [{\citenamefont {Place}\ \emph {et~al.}(2021)\citenamefont {Place},
  \citenamefont {Rodgers}, \citenamefont {Mundada}, \citenamefont {Smitham},
  \citenamefont {Fitzpatrick}, \citenamefont {Leng}, \citenamefont {Premkumar},
  \citenamefont {Bryon}, \citenamefont {Vrajitoarea}, \citenamefont {Sussman},
  \citenamefont {Cheng}, \citenamefont {Madhavan}, \citenamefont {Babla},
  \citenamefont {Le}, \citenamefont {Gang}, \citenamefont {Jäck},
  \citenamefont {Gyenis}, \citenamefont {Yao}, \citenamefont {Cava},
  \citenamefont {de~Leon},\ and\ \citenamefont {Houck}}]{Place2021}%
  \BibitemOpen
  \bibfield  {author} {\bibinfo {author} {\bibfnamefont {A.~P.~M.}\
  \bibnamefont {Place}}, \bibinfo {author} {\bibfnamefont {L.~V.~H.}\
  \bibnamefont {Rodgers}}, \bibinfo {author} {\bibfnamefont {P.}~\bibnamefont
  {Mundada}}, \bibinfo {author} {\bibfnamefont {B.~M.}\ \bibnamefont
  {Smitham}}, \bibinfo {author} {\bibfnamefont {M.}~\bibnamefont
  {Fitzpatrick}}, \bibinfo {author} {\bibfnamefont {Z.}~\bibnamefont {Leng}},
  \bibinfo {author} {\bibfnamefont {A.}~\bibnamefont {Premkumar}}, \bibinfo
  {author} {\bibfnamefont {J.}~\bibnamefont {Bryon}}, \bibinfo {author}
  {\bibfnamefont {A.}~\bibnamefont {Vrajitoarea}}, \bibinfo {author}
  {\bibfnamefont {S.}~\bibnamefont {Sussman}}, \bibinfo {author} {\bibfnamefont
  {G.}~\bibnamefont {Cheng}}, \bibinfo {author} {\bibfnamefont
  {T.}~\bibnamefont {Madhavan}}, \bibinfo {author} {\bibfnamefont {H.~K.}\
  \bibnamefont {Babla}}, \bibinfo {author} {\bibfnamefont {X.~H.}\ \bibnamefont
  {Le}}, \bibinfo {author} {\bibfnamefont {Y.}~\bibnamefont {Gang}}, \bibinfo
  {author} {\bibfnamefont {B.}~\bibnamefont {Jäck}}, \bibinfo {author}
  {\bibfnamefont {A.}~\bibnamefont {Gyenis}}, \bibinfo {author} {\bibfnamefont
  {N.}~\bibnamefont {Yao}}, \bibinfo {author} {\bibfnamefont {R.~J.}\
  \bibnamefont {Cava}}, \bibinfo {author} {\bibfnamefont {N.~P.}\ \bibnamefont
  {de~Leon}},\ and\ \bibinfo {author} {\bibfnamefont {A.~A.}\ \bibnamefont
  {Houck}},\ }\bibfield  {title} {\bibinfo {title} {New material platform for
  superconducting transmon qubits with coherence times exceeding 0.3
  milliseconds},\ }\bibfield  {journal} {\bibinfo  {journal} {Nature
  Communications}\ }\textbf {\bibinfo {volume} {12}},\ \href
  {https://doi.org/10.1038/s41467-021-22030-5} {10.1038/s41467-021-22030-5}
  (\bibinfo {year} {2021})\BibitemShut {NoStop}%
\bibitem [{\citenamefont {Reagor}\ \emph {et~al.}(2016)\citenamefont {Reagor},
  \citenamefont {Pfaff}, \citenamefont {Axline}, \citenamefont {Heeres},
  \citenamefont {Ofek}, \citenamefont {Sliwa}, \citenamefont {Holland},
  \citenamefont {Wang}, \citenamefont {Blumoff}, \citenamefont {Chou},
  \citenamefont {Hatridge}, \citenamefont {Frunzio}, \citenamefont {Devoret},
  \citenamefont {Jiang},\ and\ \citenamefont {Schoelkopf}}]{Reagor2016}%
  \BibitemOpen
  \bibfield  {author} {\bibinfo {author} {\bibfnamefont {M.}~\bibnamefont
  {Reagor}}, \bibinfo {author} {\bibfnamefont {W.}~\bibnamefont {Pfaff}},
  \bibinfo {author} {\bibfnamefont {C.}~\bibnamefont {Axline}}, \bibinfo
  {author} {\bibfnamefont {R.~W.}\ \bibnamefont {Heeres}}, \bibinfo {author}
  {\bibfnamefont {N.}~\bibnamefont {Ofek}}, \bibinfo {author} {\bibfnamefont
  {K.}~\bibnamefont {Sliwa}}, \bibinfo {author} {\bibfnamefont
  {E.}~\bibnamefont {Holland}}, \bibinfo {author} {\bibfnamefont
  {C.}~\bibnamefont {Wang}}, \bibinfo {author} {\bibfnamefont {J.}~\bibnamefont
  {Blumoff}}, \bibinfo {author} {\bibfnamefont {K.}~\bibnamefont {Chou}},
  \bibinfo {author} {\bibfnamefont {M.~J.}\ \bibnamefont {Hatridge}}, \bibinfo
  {author} {\bibfnamefont {L.}~\bibnamefont {Frunzio}}, \bibinfo {author}
  {\bibfnamefont {M.~H.}\ \bibnamefont {Devoret}}, \bibinfo {author}
  {\bibfnamefont {L.}~\bibnamefont {Jiang}},\ and\ \bibinfo {author}
  {\bibfnamefont {R.~J.}\ \bibnamefont {Schoelkopf}},\ }\bibfield  {title}
  {\bibinfo {title} {Quantum memory with millisecond coherence in circuit
  qed},\ }\href {https://doi.org/10.1103/PhysRevB.94.014506} {\bibfield
  {journal} {\bibinfo  {journal} {Phys. Rev. B}\ }\textbf {\bibinfo {volume}
  {94}},\ \bibinfo {pages} {014506} (\bibinfo {year} {2016})}\BibitemShut
  {NoStop}%
\bibitem
  [{https://www.laird.com/products/absorbers/structural-absorbers/castable-liquid-absorber/eccosorb-cr()}]{Laird}%
  \BibitemOpen
  https://www.laird.com/products/absorbers/structural-absorbers/castable-liquid-absorber/eccosorb-cr,\
  \href@noop {} {}\BibitemShut {NoStop}%
\bibitem [{\citenamefont {Ganjam}(2023)}]{GanjamThesis2023}%
  \BibitemOpen
  \bibfield  {author} {\bibinfo {author} {\bibfnamefont {S.}~\bibnamefont
  {Ganjam}},\ }\emph {\bibinfo {title} {Improving the Coherence of
  Superconducting Quantum Circuits through Loss characterization and Design
  Optimization}},\ \href@noop {} {Ph.D. thesis},\ \bibinfo  {school} {Yale
  University} (\bibinfo {year} {2023})\BibitemShut {NoStop}%
\end{thebibliography}%

\clearpage

\title{The waves-in-space Purcell effect for superconducting qubits: Supplementary information}
\maketitle

\onecolumngrid

\section{Coherence Calculation}\label{ssec:coherencecalc}

Relating the couplings of the qubit port and two potential readout port placements (WISPE and anti-WISPE) to the Rabi rates achieved when driving these ports allows us to extract the lifetimes associated with internal and external losses through the following calculation.  The transmon Hamiltonian in the presence of a drive is given by
\begin{equation}
    \Hamiltonian = \hbar\omega_q\hat{a}^\dagger\hat{a} - \frac{1}{12}E_c(\hat{a}+\hat{a}^\dagger)^4 +\hbar(\epsilon(t)+\epsilon^\star(t))(\hat{a}^\dagger+\hat{a}),
\end{equation} 
where we define the drive term $\epsilon(t) = \frac{\Omega_d}{2}e^{-i\omega_dt}$, $\Omega_d$ is the drive amplitude, and $\omega_d$ is the drive frequency. Moving this Hamiltonian to the rotating frame of the transmon then the displaced frame of the drive, we arrive at the Hamiltonian: 
\begin{equation}
    \tilde{\Hamiltonian} = -\frac{1}{12}E_c(\hat{\tilde{a}}+\hat{\tilde{a}}^\dagger-\xi-\xi^\star)^4 +\hbar (\epsilon(t)\\+\epsilon^*(t))(\hat{\tilde{a}}^\dagger+\hat{\tilde{a}}-\xi-\xi^*)- i\hbar(\dot{\xi}\hat{\tilde{a}}^\dagger +\xi\dot{\hat{\tilde{a}}}^\dagger -\dot{\xi^*}\hat{\tilde{a}}-\xi^*\dot{\hat{\tilde{a}}}).
\end{equation}
We can solve for $\xi$ by solving the differential equation 
\begin{equation}
    \dot{\xi} = -i\epsilon(t)-\left(\frac{\kappa}{2}+i\omega_p\right)\xi.
\end{equation}
Here $\xi(t) = \frac{i\Omega_{Rabi}e^{-i\omega_dt}}{\kappa+2i\Delta}$, where $\kappa$ is the decay rate due to the coupling to the 50 $\Omega$ outside environment, $\Delta$ is the frequency difference between the qubit transition frequency and the drive, and $|\xi|^2$ is the photon number \cite{Blais2021, Ganjam2024, GanjamThesis2023}. 
We can relate the photon number to $\Omega_{Rabi}$ in the following expression: 
\begin{equation}
    \bar{n} = \frac{\Omega_{Rabi}^2Q_l^2}{\omega_q^2}.
    \label{eq:photonnumber}
\end{equation}

Furthermore, if we treat the qubit as a two-level system and assume higher order transitions are minimal, we can calculate the total photon number of the system using input-output theory \cite{Ganjam2024,GanjamThesis2023}. This is photon number is the amount of photons within a cavity due to a drive. Through our initial assumptions, we correlate the same drive applied to a qubit as a photon number. Because the samples in this paper were measured in reflection, photon number can be related to input power as such: 
\begin{equation}
    \bar{n} = \frac{4}{\hbar\omega_q^2}\frac{Q_l^2}{Q_c}P_{in}.
    \label{eq:input-output}
\end{equation} 
Using Eq. \ref{eq:photonnumber} and Eq. \ref{eq:input-output}, we create a relation between the coupling Q and the input power: 
\begin{equation}
    Q_c = \frac{4P_{in}}{\hbar\Omega_{Rabi}^2}.
    \label{eq:decayrate}
\end{equation} 
This correlation is the basis of the calculations done to extract the loss rate due to the drive ports \cite{Ganjam2024,GanjamThesis2023, GeerlingsThesis}.

To capture the Purcell protection built into this system, we measured a qubit in two distinct readout scenarios to distinguish the different decay rates in the system. In the tube experiment, the readout coupler was positioned in the most and least optimal positions, which we have dubbed the WISPE and anti-WISPE ports, respectively. First, we can examine the hierarchy of losses of the worse readout coupler position as 
\begin{equation}
    \Gamma_{Total}  = \Gamma_{int} +\Gamma_{d} + \Gamma_{r}.
\end{equation}
Here, $\Gamma_{int}$ is the internal losses of the qubit, $\Gamma_{d}$ is the loss rate due to the 50 $\Omega$ coupling to the qubit drive, and $\Gamma_{r}$ is the loss rate due to the 50 $\Omega$ coupling to readout. For this anti-WISPE case, we are assuming that the lifetime of the qubit is significantly reduced by induced Purcell loss from the coupler that we can neglect the internal loss term. This assumption was justified as the lifetime of the qubit was significantly reduced between readout cases demonstrated in Figure~\ref{fig:tube}. Therefore, the expression for the losses in this least optimal coupler position is simply  
\begin{equation}
    \Gamma_{Total}  = \Gamma_{d} + \Gamma_{r}.
\end{equation}
Because of the relation between decay rate and the input power that we demonstrated in Eq.~\ref{eq:decayrate}, we can create a proportionality between $\Gamma_{d}$ and $\Gamma_{r}$  as
 \begin{equation}
    \frac{\Gamma_{d}}{\Gamma_{r}}=\frac{P_{r}}{P_{d}}.
    \label{eq:proportion}
\end{equation}
The input power needed to drive a transition from $\ket{g}-\ket{e}$ from a coupler is defined as the square of the area under the Gaussian envelope used to drive the qubit. This area can be approximated as the product of the amplitude of the pulse and length of the Gaussian envelope. In addition, there is a 13 dB attenuation difference between the readout lines and the qubit drive lines that is accounted for. Rewriting the above equation, we have $\Gamma_r=\frac{P_d}{P_r}\Gamma_d$ and using the measured lifetime of the qubit as $\Gamma_{total}$, we can solve for $\Gamma_d$. 

Having characterized the decay rate to the qubit drive port $\Gamma_d$, we can turn to the optimal readout coupler and examine the hierarchy of losses.
Here, we again use Eq. \ref{eq:proportion} measured for this configuration, the measured lifetime, and the previously calculated $\Gamma_d$ to calculate the internal loss of this qubit. With the internal loss of the qubit calculated, we can calculate the decay rate of the optimal readout coupler, $\Gamma_r$, to find the amount of Purcell protection built into the system. The data for these decay rates is represented in Fig. \ref{fig:tube}. 

Through this calculation, if we can measure the $\pi$-pulse ratios between different ports and the lifetime of the qubit from the anti-WISPE port, that is, the qubit lifetime completely dominated by Purcell loss, we can infer the Purcell protection built into our system in the WISPE configuration.

\section{3D Housings}\label{ssec:housings}

\begin{figure*}[h!]  
    \centering
	\includegraphics{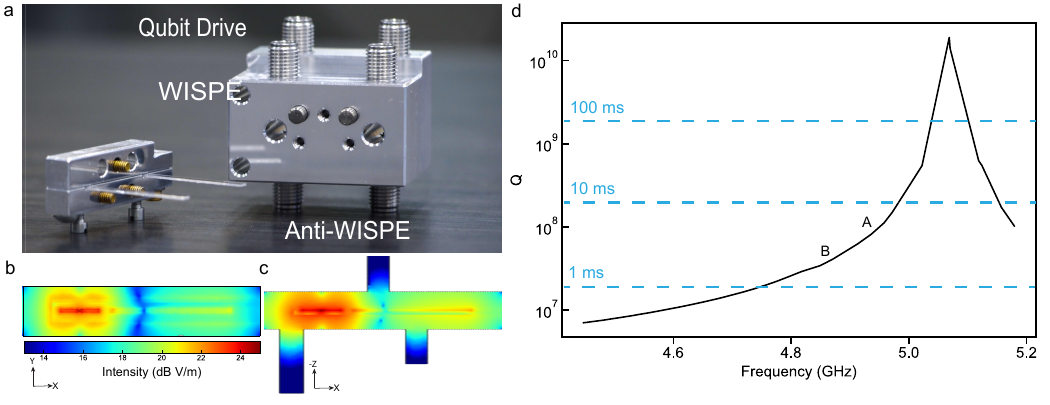}
	\caption[Purcell Effect]{\textbf{Image of qubit/cavity housing.}  a) 3D housing used in the experiment where the comparison testing was done. Two different readout configurations are possible by swapping a port between cooldowns using the same qubit. b-c) Different cuts of the qubit electric fields at its respective frequency marking the null point in the fields. d) Changing the frequency of the qubit changes the relative Purcell protection of the system. There are narrow constraints on the frequency of the qubit; however, it should be noted that even 200 MHz off from the target frequency still yields millisecond level Purcell protection with much more protection possible with a better fabricated qubit. The dashed lines correspond to simulated Purcell protection lifetimes.}
	\label{fig:panflute}
\end{figure*}
A few different Al housings were used in this project. The first one is a 4 mm diameter hole bored into a block of 6061 Al. The chips are clamped on one end and suspended inside the Al housing. A separate end cap is used to close off the other end of the tube. Coupling for the qubit drive port and readout ports were made using threaded 3D waveguide probes from Southwest Microwave (part number 1020-05SF). We can adjust the coupling of the pins to the qubit and resonator by adjusting how far the pin sticks into the tube. As the pin sticks in further into the tube, the coupling increases exponentially. For this particular system, there are 3 dedicated port holes. One of these port holes correspond to the dedicated qubit drive and the other two holes correspond to different readout port positions. Only one of the readout ports is used at a time and the other one is unscrewed from the housing. Between subsequent cooldowns, these waveguide probes can be switched out. This enables us to measure the same qubit using two different readout positions conducting our comparison experiment. A picture of this housing is shown in Fig.~\ref{fig:panflute}a.  We show the qubit electric fields of our qubit-resonator chip in Fig.~\ref{fig:panflute}b,c. 

We study how robust the WISPE point of this system is by using a finite-element simulation (HFSS) to simulate how changing the qubit frequency affects the lifetime of the qubit by changing the inductance. We note that the qubits measured and shown in Fig.~\ref{fig:panflute}d, denoted by the labels A and B, are off from the intended qubit frequency of 5.05 GHz but still demonstrate a great amount of Purcell protection as seen in Fig.~\ref{fig:tube}. This intrinsic Purcell protection is seen in other works but we have demonstrated an order of magnitude higher lifetimes in these tube geometries. 

\begin{figure}[h!]  
    \centering
	\includegraphics[]{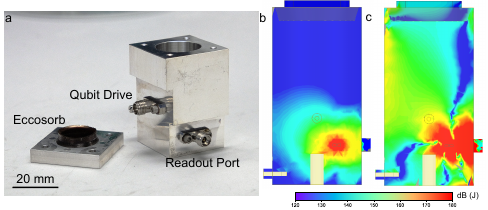}
	\caption[Purcell Effect]{\textbf{3D Post Cavity.} a) The physical 3D cavity with eccosorb used with the ports used to drive/readout the qubit are labeled. b) The qubit electric fields at its respective frequency shows that the post in the cavity creates a ``shadow'' where the qubit fields are reduced. c) Plotting $\left|\frac{\vec{E_q}\cdot \vec{E_c}}{\vec{E_c}\cdot \vec{E_c}}\right|$ shows points in space where the qubit fields are weak and where the cavity fields are strong. This guides the design of more complicated designs of where to place the readout port.}
	\label{fig:3Dpost}
\end{figure}

The 3D coaxial $\lambda  /4$ post-cavity was designed such that the center post determines the bottom cavity mode frequency. The total height of the cavity was made long enough to make sure that the higher order modes can reside in a region spatially distinct from the bottom cavity mode, where the lowest cavity mode is exponentially suppressed by the lid of the cavity. Because of this geometry, placing a lossy rf material, Eccosorb, allows us to add additional loss to the higher order modes of the cavity without affecting the lowest cavity mode shown in Fig.~\ref{fig:3Dpost}a. A sapphire chip with the qubit is suspended inside this 3D cavity. The position of the 3D chip and the readout coupler was guided by the 3D structure of the housing and regions where the qubit fields were weak but the cavity fields were strong as in Fig.~\ref{fig:3Dpost}b,c. 

For a multi-mode Purcell construction, the presence of additional loss in one of the cavity modes would shift/disrupt the destructive interference which results in a $T_1$ sweet spot seen in Fig.~\ref{fig:circuits}d. However, as noted in the main text,  we demonstrate qubit lifetime is unaffected by the addition of eccosorb into the 3D cavity, thus showing that the qubit is not affected by the multi-mode Purcell effect when below the lowest cavity mode. Additionally, Eccosorb was added to increase the $T_{2R}$ of the qubit as each of these higher order modes and the qubit's couplings to these modes provides an additional channel for decoherence. As more photons reside within the higher order cavity modes, we would expect to see photon-induced dephasing. By heavily attenuating these modes ($\kappa/2\pi =0$) or preventing photons from escaping ($\kappa/2\pi =\infty$), we could increase the $T_{2R}$ of our qubit\cite{SearsThesis}. Our technique of using Eccosorb intended on attenuating the higher order cavity modes to reduce dephasing; however, in our experiment, shown in Fig.~\ref{fig:cavity}, we note only small increase in $T_{2R}$.

\section{Higher Mode Testing}\label{ssec:highermodes}
\begin{figure*}[ht]  
    \centering
	\includegraphics[]{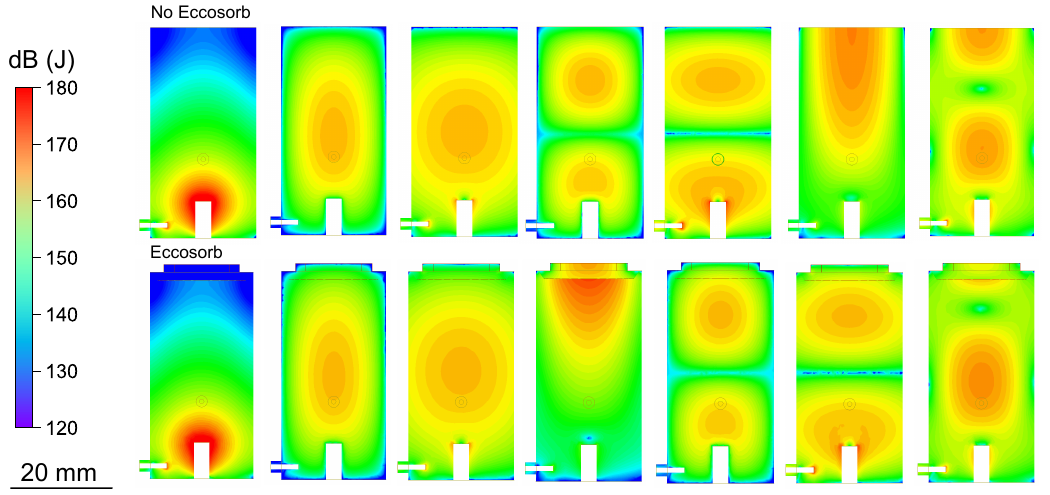}
	\caption[Eccosorb modes]{\textbf{3D Cavity modes.} The lowest seven modes of the 3D post cavity are shown for the case where there is no eccosorb and where there is 1.5 mm of eccosorb within the lid of the cavity. We see slight distortions of the electric fields in the higher order modes and becomes more apparent as the mode number increases.}
	\label{fig:3Dmode}
\end{figure*}

We show in Fig.~\ref{fig:3Dmode} the lowest seven modes of the 3D cavity with no eccosorb and with 1.5 mm of Eccosorb in the lid. We see how the lowest resonator mode is heavily attenuated closer to the lid by the natural geometry of the cavity and the higher order modes residing above the post of the cavity are perturbed slightly through the addition of Eccosorb.

\begin{figure}[h!]  
    \centering
	\includegraphics[]{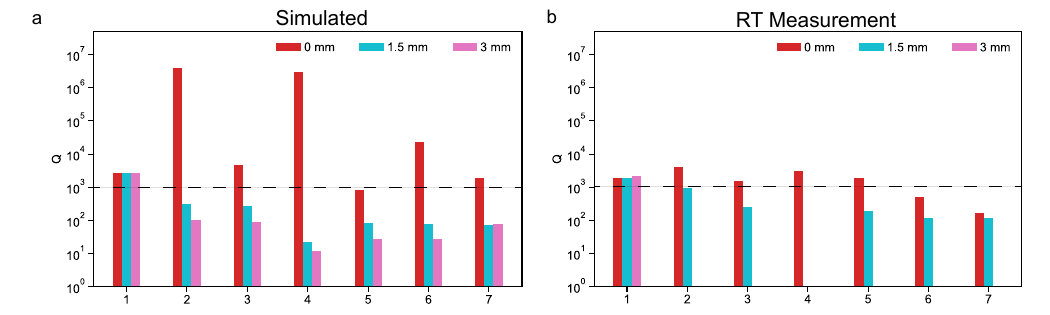}
	\caption[eccosorbloss]{\textbf{Eccosorb Loss.} Simulated and room temperature measurements of the lowest seven modes of the 3D post-cavity from 1-20 GHz. The simulated values show a significant decrease in the Q of the modes when more eccosorb is added. The RT measurements demonstrate that most of the modes vanish when higher more eccosorb is added to the lid.}
	\label{fig:barloss}
\end{figure}

The 3D housing was tested at room temperature by measuring the resonant modes using a vector network analyzer (VNA) in S11 through the readout port. Room temperature measurements were taken because the HEMT used in the experiment only provided gain between 4-8 GHz therefore providing little to no SNR on the higher order modes of the cavity. Room temperature measurements instead showed that each of the higher order modes of the cavity were broadened by the addition of Eccosorb seen in Fig. \ref{fig:barloss}. There was some slight perturbation (1-10 MHz) in the frequency of the cavity modes with the inclusion of the eccosorb, but we see a demonstrable decrease in the coupling of these modes to the readout. Because of our RT measurements, the quality factors are much lower than the simulated values. Because the external Q of the readout port was set to $10^3-10^4$, we find that modes with the 3mm of Eccosorb were so significantly broadened that we could not measure the modes above the lowest cavity.

With each lid change, we also measured a qubit in the 3D cavity to determine if the lifetimes of the qubit would be affected. By attenuating the higher order modes, we would hope to increase $T_{2R}$ by reducing the decoherence caused by the qubit coupling to higher order cavity modes. In our geometric picture of the system, the $T_1$ of the system would be unperturbed by the extinguishing other modes demonstrating how the multi-mode Purcell effect is not an effective model for simulating the system. The data for the qubit measured with differing lids is shown in Fig. \ref{fig:cavity}.

\section{Qubit Fabrication}
The qubits were fabricated using tantalum deposited on c-plane sapphire. The sapphire was first dipped in a 2:1 piranha ($H_2SO_4:H_2O_2$) solution for 20 minutes then subsequently rinsed off in 3 DI-water baths. The sapphire wafer is then transferred to the AJA sputtering system and Ta is sputtered onto the wafer at 800 C. We carefully XRD the wafer to ensure that the correct phase of Ta is grown.

With the Ta covered wafer, we do optical lithography to define the large scale features, etch away the Ta using a wet chemical process, then use e-beam lithography to define the Josephson junction. After developing the e-beam resist, the wafer is mounted into a Plassys to go through a double angle Al deposition process to make the Josephson junction using the Dolan bridge method. After probing to inspect the qubits, the wafer is sawed into 2mm by 20mm chips and mounted into the 3D Al housings. 

\begin{figure}[ht]  
    \centering
	\includegraphics[]{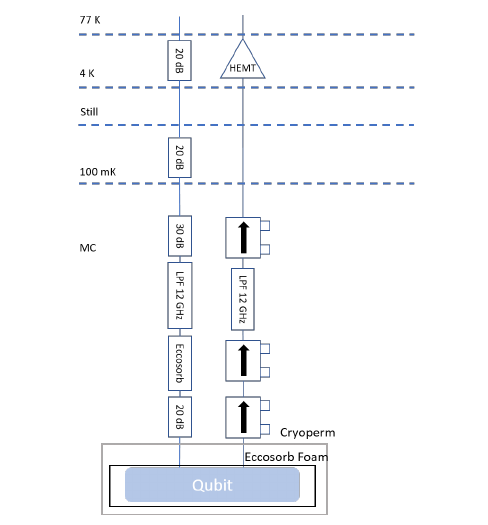}
	\caption[fridgelines]{\textbf{Fridge Diagram.} A diagram of the input and output lines of the fridge used to measure these qubit. A series of attenuators and filters were used in-line to ensure the qubit lifetime was as high as possible.}
	\label{fig:fridgelines}
\end{figure}

\end{document}